\documentclass[prb,twocolumn,nofootinbib,amsmath,amssymb,longbibliography]{revtex4-1}
\usepackage{graphicx}
\usepackage{float}
\usepackage[usenames,dvipsnames]{xcolor}
\begin{document}

\title{Depinning, Melting and Sliding of Generalized Wigner Crystals
  in Moir{\' e} Systems }
\author{
C. Reichhardt and C.~J.~O.~Reichhardt 
} 
\affiliation{
Theoretical Division and Center for Nonlinear Studies,
Los Alamos National Laboratory, Los Alamos, New Mexico 87545, USA
}

\date{\today}

\begin{abstract}

We numerically examine the depinning, sliding, and melting of commensurate and incommensurate
Wigner crystals on two-dimensional hexagonal periodic substrates near fillings of $1/3,1/2$ and $2/3$ to model the dynamics of generalized Wigner crystals
in moir{\' e} heterostructures. At low temperatures where thermal
fluctuations are irrelevant,
for commensurate fillings
we find a strongly pinned state that depins into a sliding crystal, while at incommensurate fillings, the depinning threshold is strongly reduced. For fillings below $1/2$, the depinning occurs in a two-step process. Above the first
depinning threshold, there is
an extended range of drives where the
conduction occurs via the sliding of anti-kinks along the charge stripes,
followed by a second threshold where all of the charges begin to slide.
At finite temperatures, the external driving reduces the effective melting temperature at commensurate fillings and enhances creep at incommensurate fillings. We show that depinning into different sliding states, such as moving fluids or moving crystals, produces nonlinear features in the transport curves. We also show that transport is asymmetric on either side of a commensurate filling due to the different dynamics for interstitials versus holes in the commensurate structure. A variety of sliding states should be accessible for generalized Wigner crystals at finite temperatures even for low drives.
If experiments can realize
sliding dynamics in moir{\' e} systems, it would open a new
class of commensurate and incommensurate phases for study.
Additionally, the sliding
dynamics and anti-kink flow could provide
new functionalities for charge transport in moir{\' e} heterostructures.
\end{abstract}

\maketitle

\section{Introduction}

Electrons in two-dimensional systems are expected to form a Wigner crystal
at low densities when the Coulomb energy is much stronger than then the kinetic
energy
\cite{Wigner34,Shayegan22}.
Evidence
for Wigner crystal formation has been
obtained via transport \cite{Goldman90,Williams91,Jiang91,Yoon99,Ikegami09,Knighton18,Falson22}, noise \cite{Brussarski18,Madathil23,Yu24},
pinning resonance \cite{Ye02,Chen06,Andrei88,Zhao23,Freeman24},
and a variety of other measures \cite{Tiemann14,Jang17,Smolenski21,Zhou21}.
Very recently,
Wigner crystal states have been imaged directly in graphene heterostructures \cite{Tsui24}.
One of the characteristic behaviors of Wigner crystals
is that they form
insulating or pinned states
that have sharp conduction thresholds
and nonlinear transport curves along with
enhanced noise near the depinning threshold
\cite{Williams91,Cha94a,Cooper99,Reichhardt01,Piacente05,Brussarski18,Reichhardt22,Reichhardt23,Madathil23,Yu24}.
Such behaviors could be tested in
a new class of charge-ordered systems
called generalized Wigner crystals
that has been realized
in moir{\' e} heterostructures where the charges interact with
periodic triangular lattices.
In these systems, the charge density or filling can be controlled,
and for fillings of $1/3$, $1/2$, $2/3$,
or other higher-order fillings, 
commensurate charge crystals form
that act as insulating states with long-range order
\cite{Regan20,Xu20,Li21,Jin21,Padhi21,Huang21,Ung23,Chen23,Tan23,Li23,Li24}.
Computational studies of the melting of generalized Wigner crystal states
under increasing temperature show that
commensurate fillings have a much higher melting temperature than
incommensurate fillings \cite{Matty22}.
One of the next steps is to
examine the transport and dynamics of generalized Wigner crystals
in order to determine whether
they can exhibit depinning phenomena similar
to what has been observed in other Wigner crystal systems.

The sliding dynamics of
generalized Wigner crystals should be a strong function of filling,
with different dynamics
appearing for commensurate versus incommensurate states.
Sliding states may occur at incommensurate fillings,
in the presence of finite temperatures, or under strong external drives.
The generalized Wigner crystal 
represents a new class of commensurate-incommensurate system that
could exhibit sliding behaviors.
Previous studies on the 
depinning of superconducting vortices
\cite{Baert95,Harada96,Reichhardt97,Reichhardt01a,Gutierrez09},
colloidal particles \cite{Bohlein12,Vanossi12,McDermott13a,Juniper15},
frictional systems \cite{Bak82,Vanossi13},
and other particle-like systems on periodic substrates \cite{Reichhardt17}
have shown that the threshold for sliding is strongly reduced at incommensurate filings due to the formation of kinks or anti-kinks.
In generalized Wigner crystals, where long-range interactions are relevant,
such kinks could be much more mobile than in the systems with
shorter-range interactions considered previously.
If sliding states or kink motion
for generalized Wigner crystals could be realized,
this could pave the way for
new functionalities for moir{\' e} systems,
such as discrete charge transport,
switching, and other novel charge transport-based devices.

In this work, we examine the depinning, sliding, and melting of
generalized Wigner crystals
coupled to a hexagonal substrate at
commensurate fillings of $\nu = 1/3$, $1/2$ and $2/3$,
as well as at nearby incommensurate fillings formed by doping with
extra particles or holes.
The charges form a triangular lattice
at $\nu = 1/3$, a stripe lattice
at $\nu=1/2$, and
a honeycomb lattice at $\nu=2/3$,
as seen in experiment \cite{Li21} and in
previous simulations \cite{Matty22,Ung23}.
This model includes the hexagonal periodic potential
expected to be present in the moir{\'e} heterostructures where
long-range charge order has been observed.
At zero or low temperatures under an external bias,
there is a sharp conduction threshold for
charge sliding at commensurate filings,
while at incommensurate fillings, the depinning thresholds
are reduced and also become less sharp.
We show that if the stripe lattice state at $\nu=1/2$ is doped to
slightly lower fillings
with holes,
the depinning threshold
is reduced by a factor of ten and
initial motion occurs by the one-dimensional
transport of anti-kinks or holes along the stripes.
There is a second depinning threshold at higher drives where
all of the charges depin, giving a two-step
depinning process that is visible in the transport curves.
If the $\nu=1/2$ state is doped to slightly higher fillings with
extra particles,
the depinning threshold is reduced but not as much as for
hole doping, since the additional dopant charges are located
between the stripes and soliton motion does not occur.
At a filling of $\nu=1/3$,
above the melting transition temperature $T_m$ the
charge positions disorder due to thermally induced
hopping from site to site.
The melting temperature diminishes as the external
driving is increased, and the particles depin into a moving liquid state;
however, at low temperatures, the melting behavior disappears and the
particles
depin into a sliding hexagonal crystal.
There is also an intermediate regime where
the system is in a fluid state with high velocity just above
the depinning transition, but reorders at higher drives
into a moving crystal state with a lower velocity.
The dynamical formation of the moving crystal
produces clear signatures in the transport and differential resistance curves.

At low but finite driving forces, both
the commensurate and incommensurate states
remain pinned at low temperatures. When the
temperature is increased, the incommensurate states
depin at much lower temperatures than the commensurate states
due to the thermally induced
hopping of kinks or interstitials.
Even for the commensurate states,
the depinning temperatures may remain
well below the melting temperature for zero
bias or zero driving force.
As a result,
the sliding states of both the
commensurate and incommensurate states should be
accessible experimentally at finite temperatures.
The $T=0$ depinning threshold is strongly
reduced near $\nu = 0.42$, where there is
a spatial phase separation between pinned and moving regions.
At $\nu=2/3$, where the system forms a pinned honeycomb lattice,
there is a discontinuous 
low temperature depinning transition
to a floating hexagonal lattice,
which slides at a slightly higher velocity then
that of the nearby incommensurate dopings.
This behavior is the opposite of what is found
for the $\nu=1/3$ and $\nu=1/2$ fillings,
where the commensurate flow has a lower velocity than the
incommensurate flow.
There is a region of soliton motion in the depinning process for
fillings below $\nu=2/3$, but except for the $\nu=1/2$ filling,
the soliton transport occurs via a two-dimensional zig-zag
anti-kink motion.

Our results suggest that generalized Wigner crystals
with long-range charge interactions
can exhibit
a wide variety of commensurate and incommensurate sliding phases, and
that
the sliding of the
incommensurate phases can be induced easily
at finite temperatures.
The realization of sliding or soliton motion for charges in
moir{\' e} heterostructures would
open up the possibility for various
new functionalities and devices based on Wigner crystal transport.

\section{Model}

We consider a two-dimensional system of
pointlike interacting charges on a hexagonal 
substrate, representing
generalized Wigner crystals in moir{\' e} heterostructures.
We place $N$ charges on a substrate containing $N_{p}$ potential minima,
giving a filling factor of $\nu \equiv N/N_{p}$.
The charges can also be subjected to
external driving forces and thermal fluctuations.
The overdamped dynamics of charge $i$ are determined by the
equation of motion

\begin{equation}
 \eta \frac{d {\bf R}_{i}}{dt} =
-\sum^{N}_{j \neq i} \nabla V(R_{ij})  + {\bf F}^{s}_{i} +  {\bf F}_{D} + {\bf F}^{T} \ .
\end{equation}

This overdamped approximation, which is expected to be valid when
the mass is small or the damping is high,
was used in previous studies of
Wigner crystal depinning and sliding on random substrates
\cite{Reichhardt01,Reichhardt04,Reichhardt22,Reichhardt23},
and gave transport behaviors similar to what was observed
in experiment \cite{Brussarski18,Madathil23,Yu24}. 
Here $\eta$ is the damping constant,
$V(R_{ij}) = q/R_{ij}$ is the Coulomb interaction,
and the distance between charges $i$ and $j$ at positions ${\bf R}_i$ and
${\bf R}_j$ is
$R_{ij}=|{\bf R}_i-{\bf R}_j|$.
To treat the long range interactions,
we employ a Lekner real-space summation technique
\cite{Lekner91,GronbechJensen97a}.

The hexagonal periodic potential is modeled with three sinusoidal
functions,
\begin{equation}
  {\bf F}_i^s = \sum^{3}_{k=1}F_p\sin(2\pi b_{i}/a_{0})[\cos(\theta_{k}){\hat {\bf x}}
    - \sin(\theta_{k}){\hat{\bf y}}].   
\end{equation}
Here, $F_p$ is the strength of the substrate
which we fix to $F_p=0.0875$,
$a_0$ is the substrate lattice constant which we fix to $a_0=2.309$,
$b_i = x_i\cos(\theta_{k}) - y_i\sin(\theta_{k}) + a_{0}/2$,
$\theta_{1} = \pi/6$, $\theta_{2} = \pi/2$,
$\theta_{3} = 5\pi/6$, $x_i={\bf R}_i \cdot {\hat{\bf x}}$,
and $y_i={\bf R}_i \cdot {\hat{\bf y}}$.
The term ${\bf F}^T$ arises from thermal fluctuations
represented by Langevin kicks
with the properties
$\langle {\bf F}_i^{T}\rangle = 0$
and $\langle {\bf F}^T_i(t){\bf F}_j^{T}(t^\prime)\rangle = 2k_BT\delta_{ij}\delta(t-t^\prime)$.
The particle positions are determined by placing the particles at the locations
of the nearest commensurate filling and then removing or adding particles
at random in order to dope away from this filling as desired.
After the system has been filled with particles,
we apply a driving force of ${\bf F}_{D}=F_D {\bf \hat{x}}$
which could be produced by applying a voltage
to the charges,
and we measure the time-averaged particle velocity in the direction of
the drive,
$\langle V\rangle = \sum^{N}_i{\bf v}_i\cdot {\hat {\bf x}}$.
We use a simulation time step of size
$\Delta \tau = 0.002$.
The drive is increased in
increments of $\Delta F_D=0.025$ or smaller, and we hold the drive
fixed for $5\times 10^7$ simulation time steps at each increment,
which is long enough to average over any transients in the
dynamics for the parameters we consider in this work.

\begin{figure}
  \includegraphics[width=\columnwidth,trim={20 0 90 0},clip]{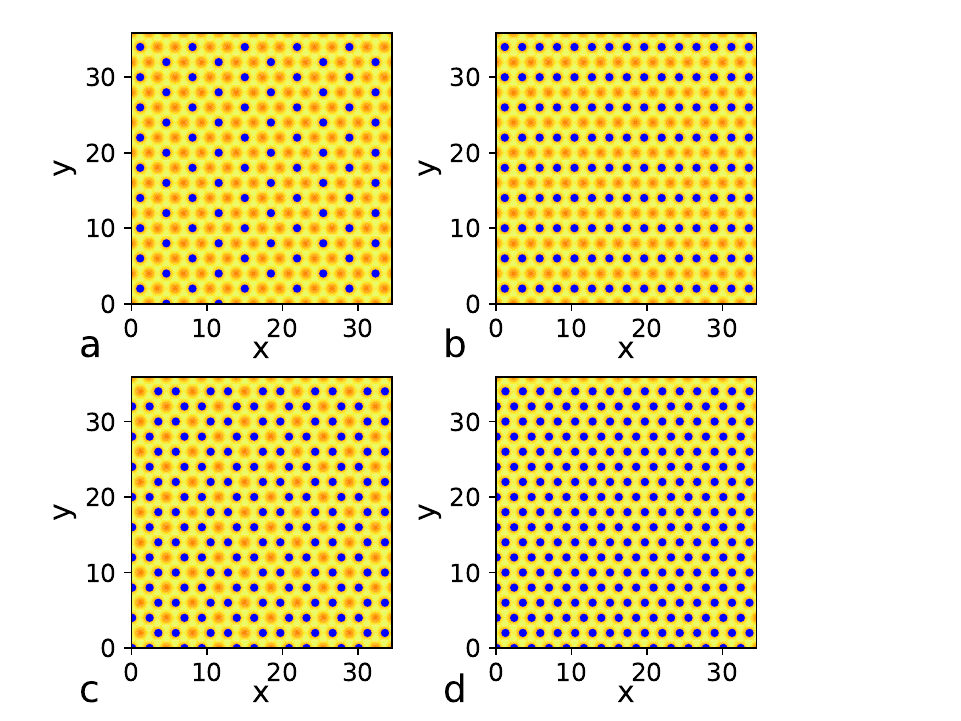}  
\caption{
The particle positions (blue circles) and centers of the
hexagonal substrate potential minima (orange circles)
for
(a) the hexagonal ordering at $\nu = 1/3$,
(b) a stripe crystal at $\nu = 1/2$,
(c) a honeycomb lattice at $\nu= 2/3$, and
(d) a hexagonal lattice at $\nu = 1.0$.
	} 
\label{fig:1}
\end{figure}

\section{1/3 Filling}

In Fig.~\ref{fig:1}(a), we show the charge-ordered states
at a filling
of $\nu = 1/3$, where the particles form a hexagonal lattice
that matches the symmetry of a subset of the substrate minima.
Figure~\ref{fig:1}(b) shows the formation of a stripe crystal state
at $\nu = 1/2$, Fig.~\ref{fig:1}(c) illustrates a honeycomb lattice
at $\nu=2/3$, and Fig.~\ref{fig:1}(d)
shows a commensurate hexagonal lattice at $\nu=1.0$.
The ordered states in Fig.~\ref{fig:1}
are the same ones that have been observed or that are expected
to occur in
generalized Wigner crystal systems
\cite{Regan20,Xu20,Li21,Padhi21,Huang21,Matty22,Ung23}.
We initially focus on fillings in the range $0.27 < \nu <  0.407$
centered around $\nu = 1/3$.

\begin{figure}
\includegraphics[width=\columnwidth]{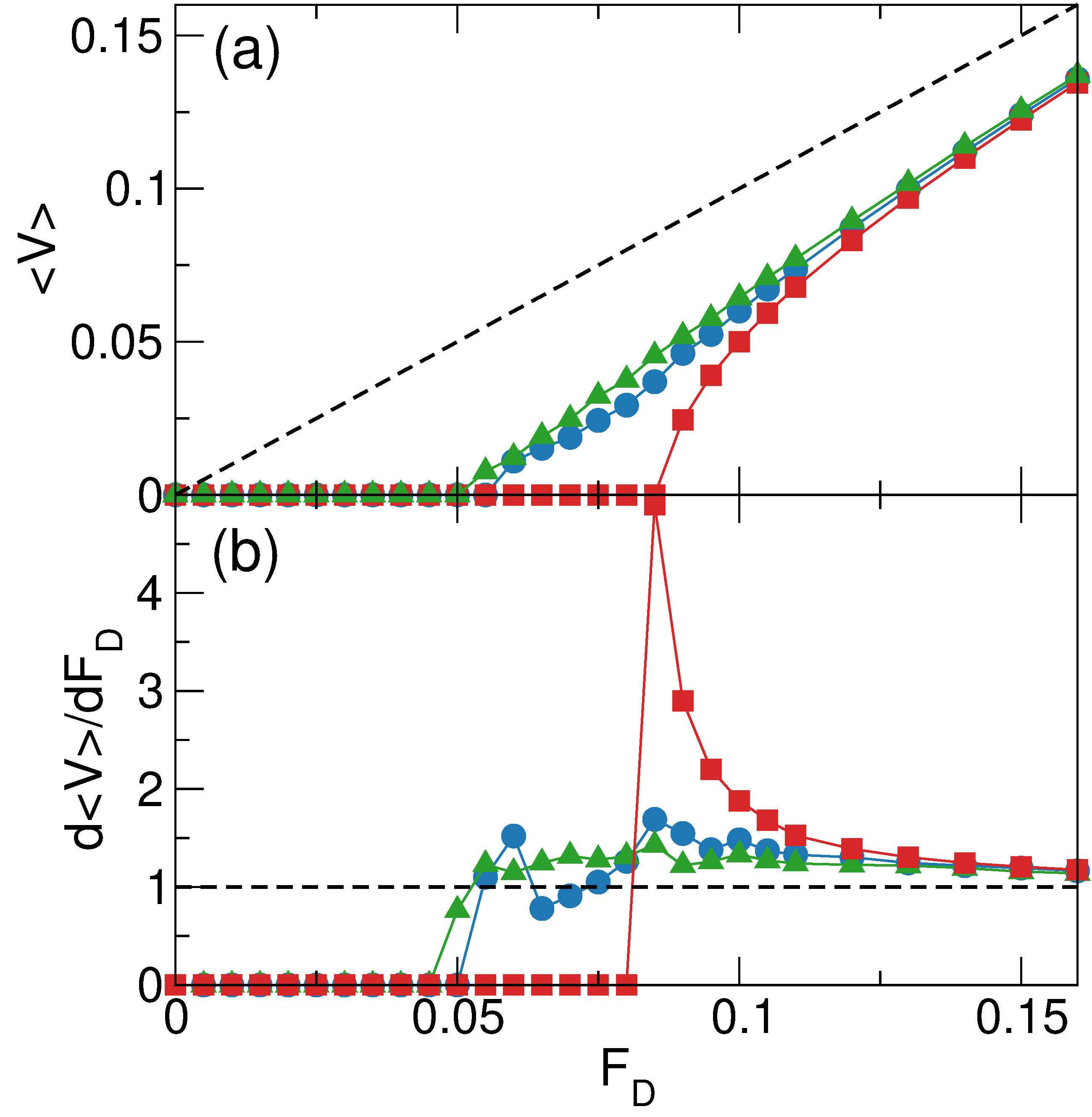}
\caption{
(a) The average velocity per particle $\langle V\rangle$
vs $F_{D}$ at $T=0$
for $\nu = 1/3$ (red squares), $\nu = 0.31$ (blue circles),
and $\nu = 0.36$ (green triangles). The dashed line is the
expected velocity in the absence of a substrate.
The $\nu = 1/3$ state depins at $F_D=F_p$, where $F_p$ is the
substrate strength.
(b) The corresponding $d\langle V\rangle/dF_{D}$ curves vs $F_{D}$. 
}
\label{fig:2}
\end{figure}

\begin{figure}
  \includegraphics[width=\columnwidth,trim={20 70 90 90},clip]{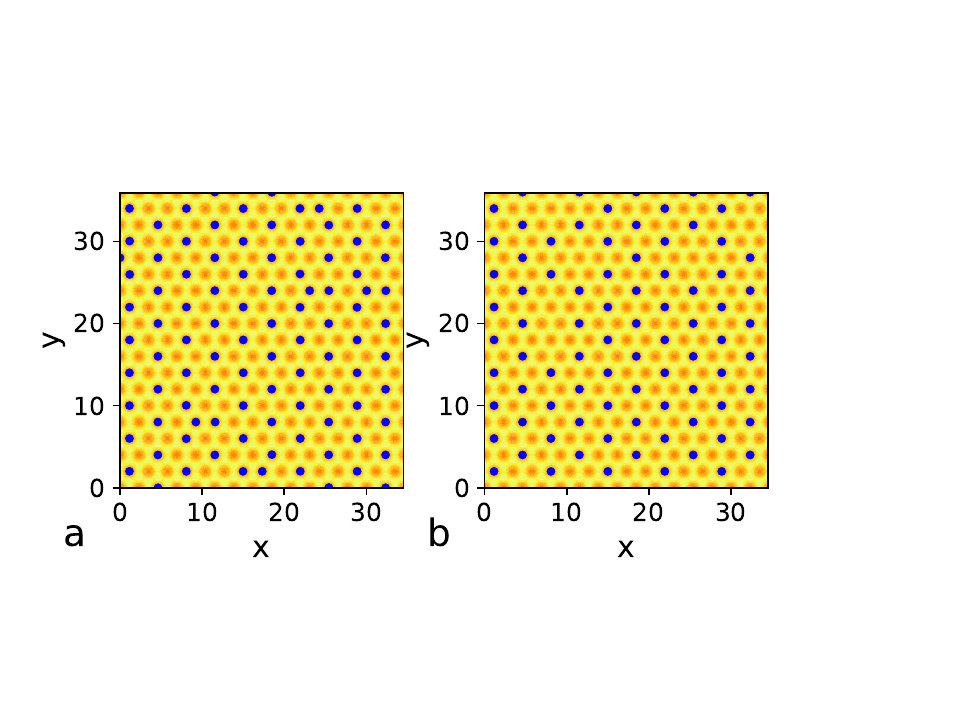}  
\caption{
Particle positions (blue circles) and centers of the hexagonal substrate
potential minima (orange circles) for the system from
Fig.~\ref{fig:2} at $T=0$ and $F_D=0$.
(a) Some dimer states are present at $\nu = 0.36$.
(b) The configuration at $\nu = 0.31$ consists of
vacancies in the $\nu = 1/3$ state.
}
\label{fig:3}
\end{figure}

\begin{figure}
  \includegraphics[width=\columnwidth,trim={20 0 90 0},clip]{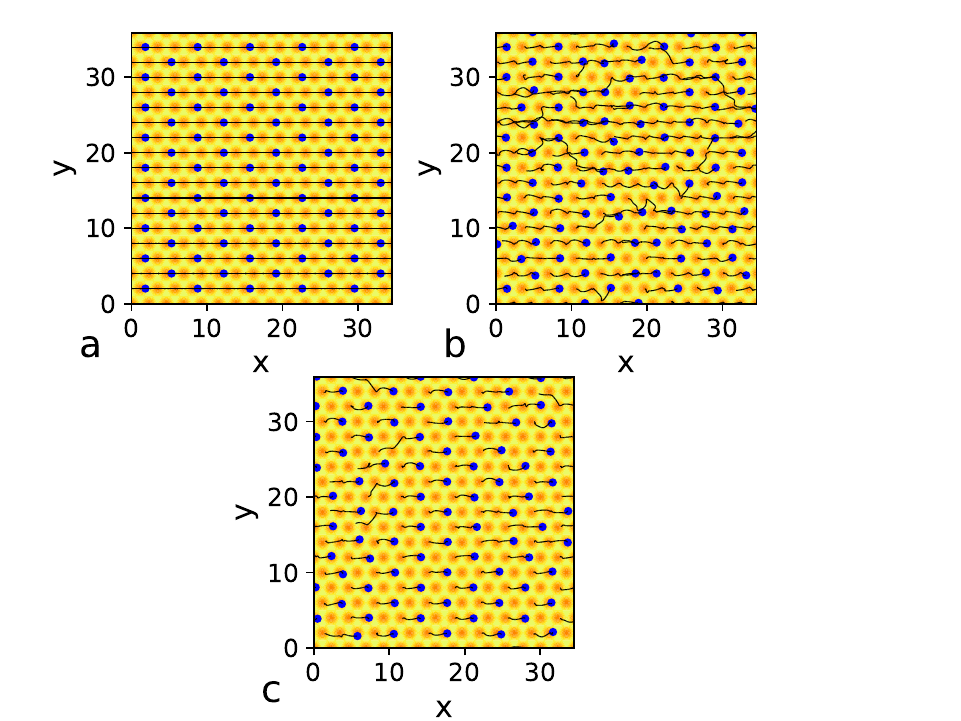}
\caption{
The particle locations (blue circles), particle trajectories (lines),
and centers of the hexagonal substrate potential minima (orange circles)
for the system in Fig.~\ref{fig:2} at $T=0$
just above depinning for
(a) $\nu = 1/3$, (b) $\nu = 0.36$, and (c) $\nu = 0.31$.
The system depins
elastically into a moving lattice at $\nu = 1/3$,
and at the incommensurate fillings 
the flow is more disordered.
}
\label{fig:4}
\end{figure}

In Fig.~\ref{fig:2}(a) we plot the velocity $\langle V\rangle$
versus drive $F_{D}$ curves
at $T=0$ for
$\nu = 0.31$, $1/3$, and $0.36$.
The dashed line indicates the expected
shape of the velocity versus force curve in the absence of a substrate.
Figure~\ref{fig:3}(a,b) shows images of the
incommensurate charge-ordered states.
In Fig.~\ref{fig:3}(a) at
$\nu = 0.36$, some dimer states appear
due to the doping of excess particles into the $\nu = 1/3$
hexagonal charge lattice,
while for $\nu = 0.31$ in Fig.~\ref{fig:3}(b),
vacancies appear
in the $\nu = 1/3$ lattice.
As shown in Fig.~\ref{fig:2}(a),
the depinning threshold $F_{c}$ is highest when $\nu = 1/3$,
reaching a value of nearly $F_c=F_p$,
and it is reduced at the incommensurate fillings.
At the $\nu=1/3$ filling, the long range ordering of the system causes
interactions between neighboring charges to cancel exactly,
so the depinning threshold is determined solely by the strength of the
substrate.
At the incommensurate fillings, higher energy
excitations appear in the form of kinks or
anti-kinks that can depin individually
or cause patches of the system to depin at much
lower drives than the depinning threshold of the kink-free lattice.
The velocity-force curve
at the $\nu=1/3$ filling behaves as
$\langle V\rangle \propto (F_{D} - F_{c})^\beta$ with
$\beta \approx 2/3$, close to the expected form for an
elastic depinning process in which
the particles maintain their same neighbors 
\cite{Fisher98,Reichhardt17}.
The pinned hexagonal lattice at $\nu=1/3$ depins
directly into a moving hexagonal lattice with the same symmetry,
as shown in Fig.~\ref{fig:4}(a), where the lines highlight the
trajectories of the charges.
The depinning threshold is reduced for $\nu = 0.31$
and the
velocity-force scaling relation that appeared at $\nu = 1/3$
is lost,
while for $\nu = 0.36$, the threshold is reduced slightly further
due to the presence of
the dimer states shown in Fig.~\ref{fig:3}(a).
Since the particles composing
the dimers are close together, the dimers are more mobile than the vacancies
that appear for $\nu=0.31$.
The particle trajectories above depinning are more disordered 
for the incommensurate fillings,
as shown in Fig.~\ref{fig:4}(b,c) for $\nu = 0.36$ and
$\nu=0.31$, respectively.
Here, there is occasional motion of the particles in the direction
transverse to the drive,
and some particles remain
pinned for a time while other particles move around them,
resulting in plastic flow.

Figure~\ref{fig:2}(b) shows the differential velocity curves,
$d\langle V\rangle/dF_{D}$ versus $F_{D}$, for the
system from Fig.~\ref{fig:2}(a).
At $\nu=1/3$, $d\langle V\rangle/dF_D$ exhibits
a pronounced single sharp peak.
For the incommensurate fillings, $d\langle V\rangle/dF_D$ is
more rounded and shows signs of a multistep 
depinning process.
This is the most clearly visible
for the $\nu = 0.31$ filling, where the first peak in
$d\langle V\rangle/dF_D$ corresponds to the depinning of the vacancies,
and the second peak falls near
$F_{D}/F_{p} = 1.0$ when all of the particles have depinned.
In the absence of a substrate, the velocity-force curve is linear and
$d\langle V\rangle/dF_D=1.0$ for all $F_D$.
In the presence of a substrate,
at high drives all of the $d\langle V\rangle/dF_D$ curves
gradually approach the substrate-free value.

\begin{figure}
\includegraphics[width=\columnwidth]{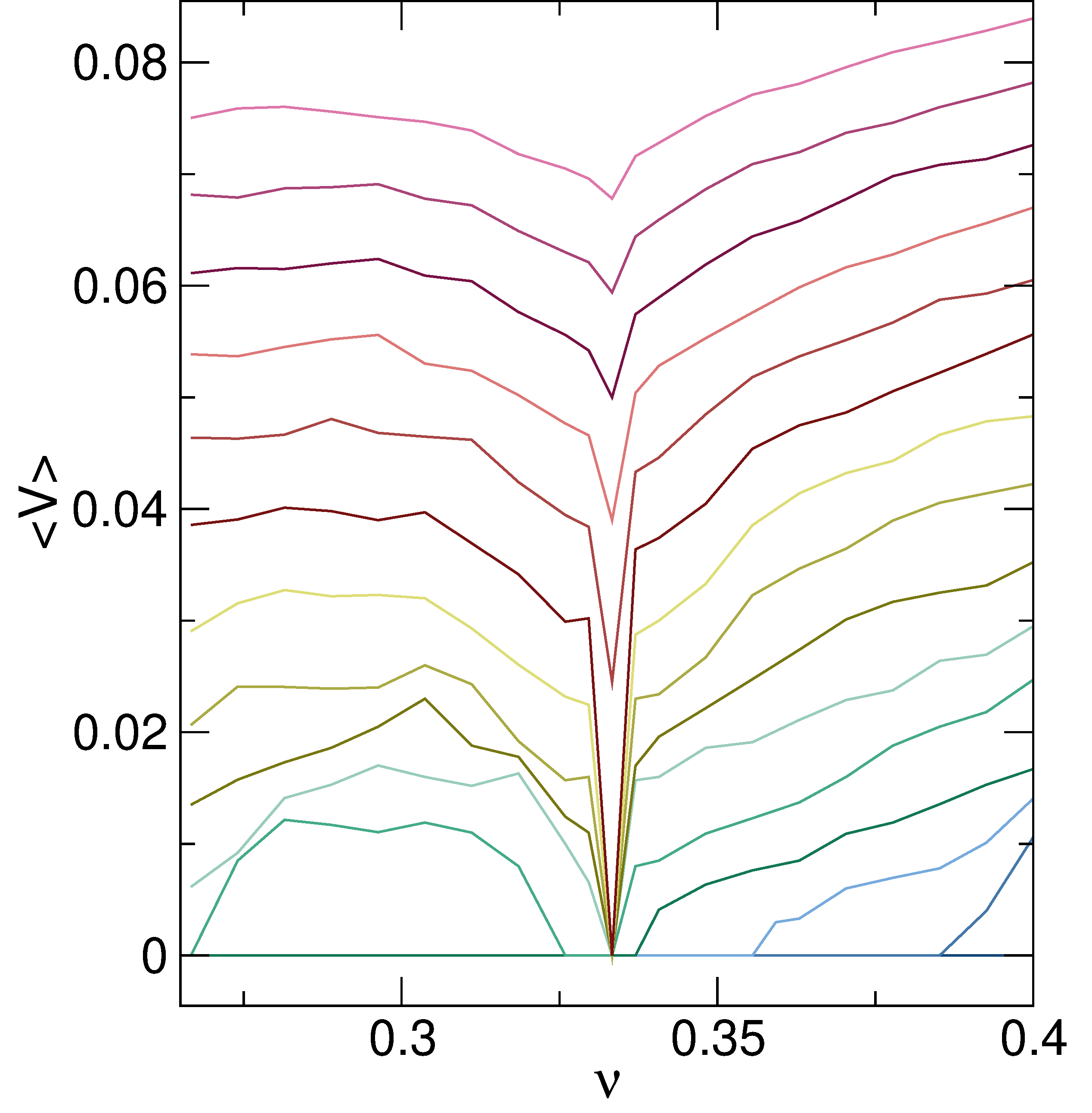}
\caption{$\langle V\rangle$ vs $\nu$ over the range
$0.27 < \nu <  0.407$ for the system from Fig.~\ref{fig:2}
at $T=0$ for
$F_D=0.035$, 0.04, 0.045, 0.05, 0.055, 0.06, 0.065, 0.07, 0.075, 0.08,
0.085, 0.09, 0.095, 0.1, 0.105, and 0.11, from bottom to top.
The system remains pinned up to $F_{D} = 0.0875 = F_{p}$
at $\nu = 1/3$, and for higher drives there is a persistent
minimum in the velocity for this value of $\nu$.
For any given value of $F_D$,
the velocities for $\nu > 1/3$ are generally higher than for
$\nu < 1/3$.
}
\label{fig:5a}
\end{figure}

\begin{figure}
\includegraphics[width=\columnwidth]{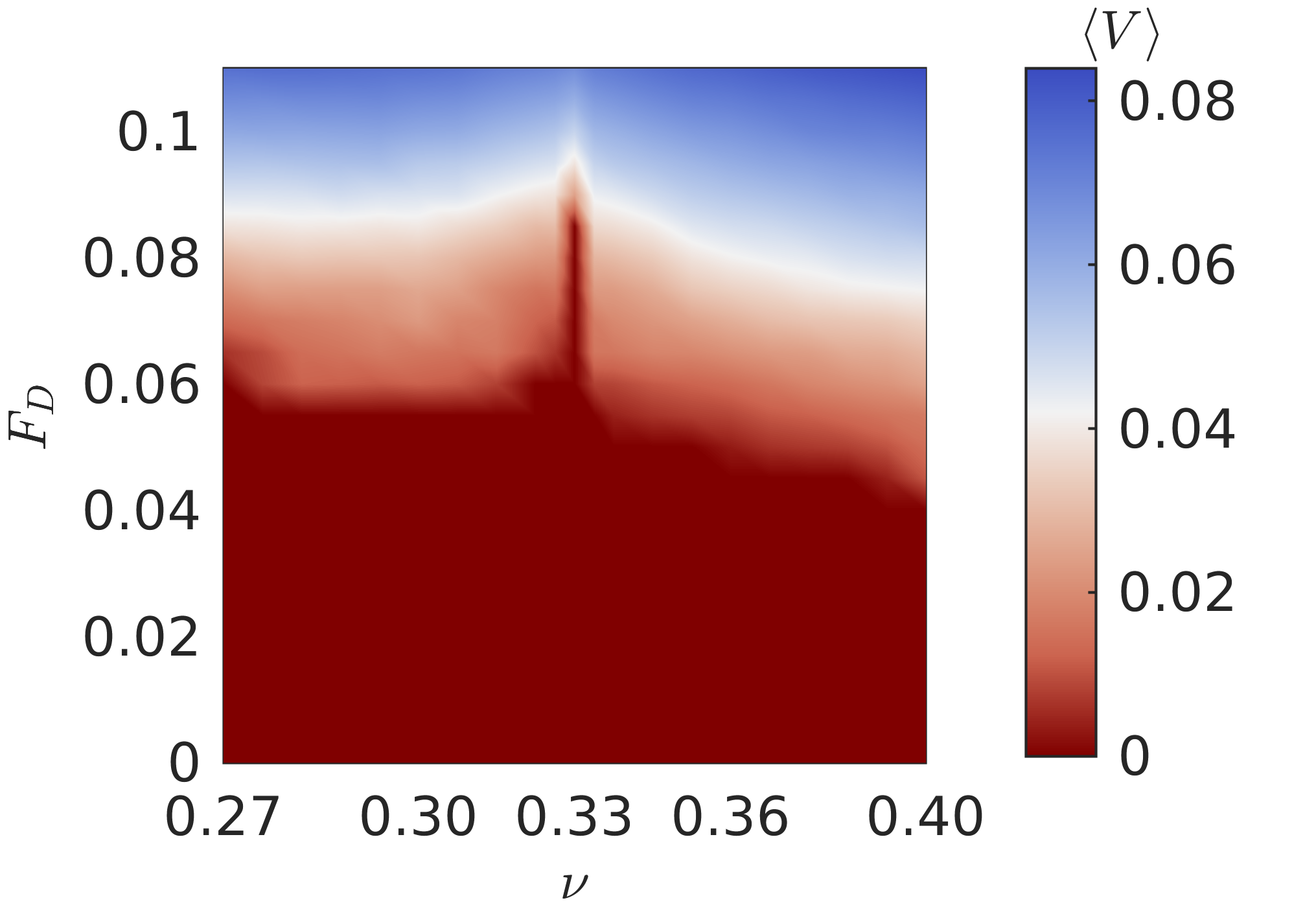}
\caption{A heat map of the velocity $\langle V\rangle$ as a function
of $F_D$ versus $\nu$ constructed using the data from Fig.~\ref{fig:5a}
in a system with $T=0$.
The pinned state and reduced velocity at $\nu = 1/3$ are clearly visible.
}
\label{fig:5b}
\end{figure}

In Fig.~\ref{fig:5a} we plot
$\langle V\rangle$ versus $\nu$ over the range
$0.27 < \nu <  0.407$ for the system from Fig.~\ref{fig:2}
at drives spanning $F_D=0.035$ to $F_D=0.11$
in increments of $\delta F_{D} = 0.005$.
When $F_D\leq 0.04$, all of the fillings are pinned and $\langle V\rangle=0$
everywhere, while
for $0.045 \leq F_D < 0.06$, only fillings with $\nu \leq 1/3$ are pinned.
For $0.06 \leq F_{D} < 0.0875 = F_p$,
$\langle V\rangle = 0$ only when $\nu = 1/3$.
We find that the velocity is generally greater for $\nu>1/3$ than for
$\nu<1/3$ at a given value of $F_D$ due to the greater mobility of
the dimer-like states compared to vacancies.
For $F_{D} > 0.0875$, there is a persistent
velocity dip at $\nu = 1/3$,
indicating that the ordered state
remains more strongly coupled to the substrate even while sliding.
In Fig.~\ref{fig:5b} we plot
a heat map of the velocity as a function of $F_D$ versus $\nu$ for
the system in Fig.~\ref{fig:5a},
highlighting the reduced velocity at $\nu = 1/3$ and the
asymmetry in the velocities above and below $\nu = 1/3$.

\section{1/2 filling}

\begin{figure}
\includegraphics[width=\columnwidth]{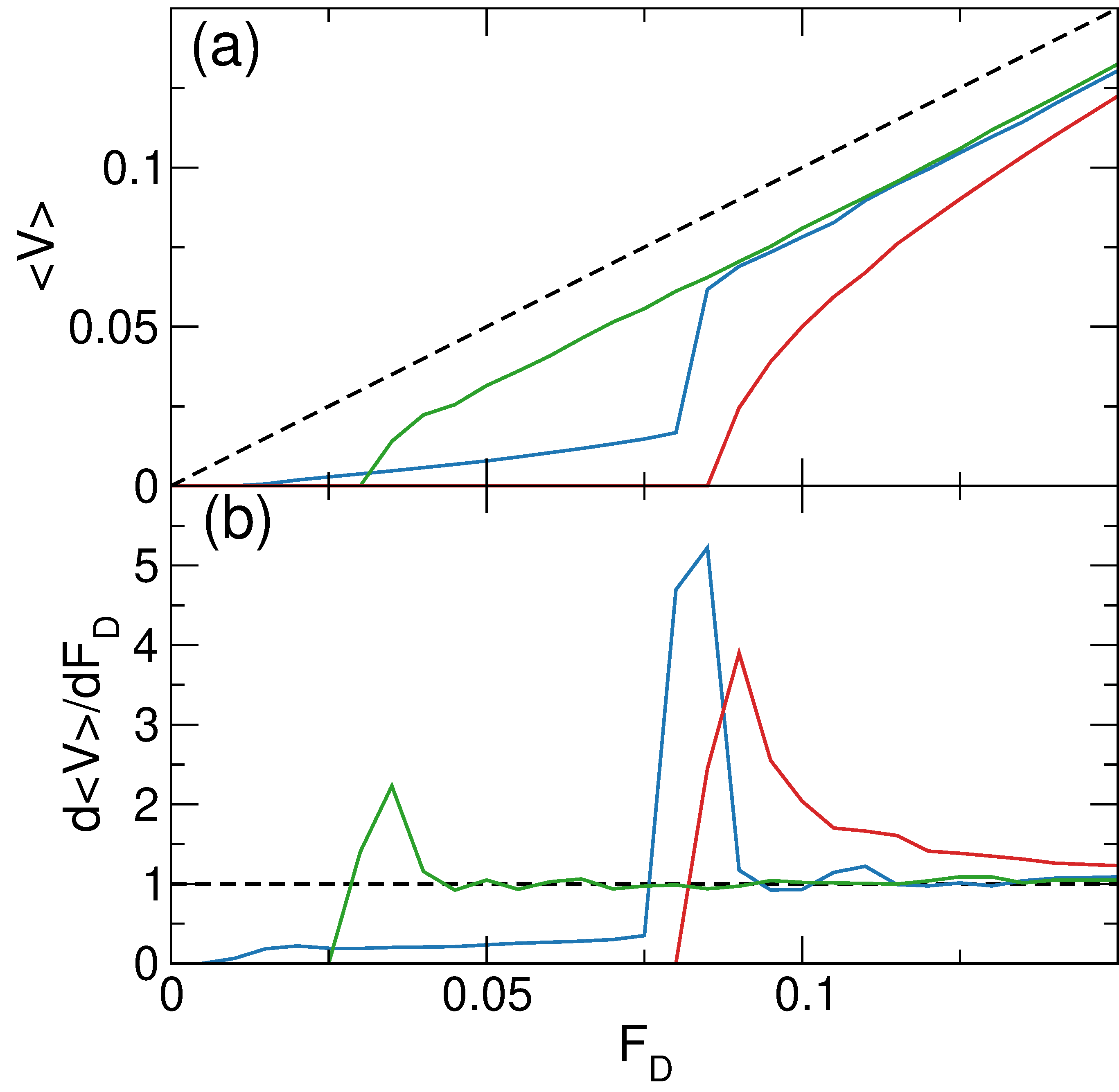}
\caption{
(a) $\langle V\rangle$ vs $F_{D}$ for a sample with $T=0$
at the $\nu = 1/2$ stripe state from Fig.~\ref{fig:1}(b) (red),
$\nu = 0.478$ (blue) and $\nu = 0.525$ (green);
the dashed line is the expected velocity in the absence of a substrate.
There is a clear two-step depinning process for $\nu = 0.478$
and a single depinning process at $\nu = 1/2$.
(b) The corresponding $d\langle V\rangle/dF_{D}$ vs $F_D$ curves. }       
\label{fig:6}
\end{figure}

We next consider the depinning and transport at the
$\nu=1/2$ filling, where a stripe crystal forms as
illustrated in Fig.~\ref{fig:1}(b).
In Fig.~\ref{fig:6}(a), we plot
$\langle V\rangle$ versus $F_{D}$
at $\nu = 1/2$, $\nu = 0.478$, and $\nu = 0.525$, along with a
dashed line indicating the expected velocity signature in the absence
of a substrate.
We find the highest possible depinning threshold of $F_c=F_p$ at
$\nu = 1/2$,
since the symmetry of the stripe state causes
all of the particle-particle interactions to cancel.
The depinning threshold for $\nu = 0.478$ is nearly
ten times lower than $F_c$ for the $\nu = 1/2$ filling,
and there is an extended
region from $0.01 < F_{D} < 0.075$ where a finite but reduced
flow occurs for $\nu=0.478$,
followed by a second depinning transition
near $F_{D} = 0.75$ that is accompanied by a jump up in the velocity.
For $\nu = 0.525$, the depinning threshold is
four times smaller than $F_c$ for $\nu = 1/2$.
Despite the fact that $F_c$
for $\nu = 0.525$ is
three times higher than
$F_c$ at $\nu = 0.478$,
the velocity in the moving state is much
larger for $\nu = 0.525$ than for $\nu=0.478$
until $F_{D} = 0.075$,
above which the velocity is almost the same for both of the two
incommensurate fillings.
In the $d\langle V\rangle/dF_D$ versus $F_D$ curves
plotted in Fig.~\ref{fig:6}(b),
a single peak appears for $\nu = 1/2$,
while the $\nu = 0.478$ filling shows a small peak
near the initial depinning transition and a sharp peak near the second
depinning transition at $F_{D} = 0.075$.
For $\nu = 0.525$, there is
a small peak in $d\langle V\rangle/dF_D$ near depinning,
and at high drives, all of the curves approach a linear velocity-force
relation with $d\langle V\rangle/dF_D \approx 1$.

\begin{figure}
    \includegraphics[width=\columnwidth,trim={20 0 90 0},clip]{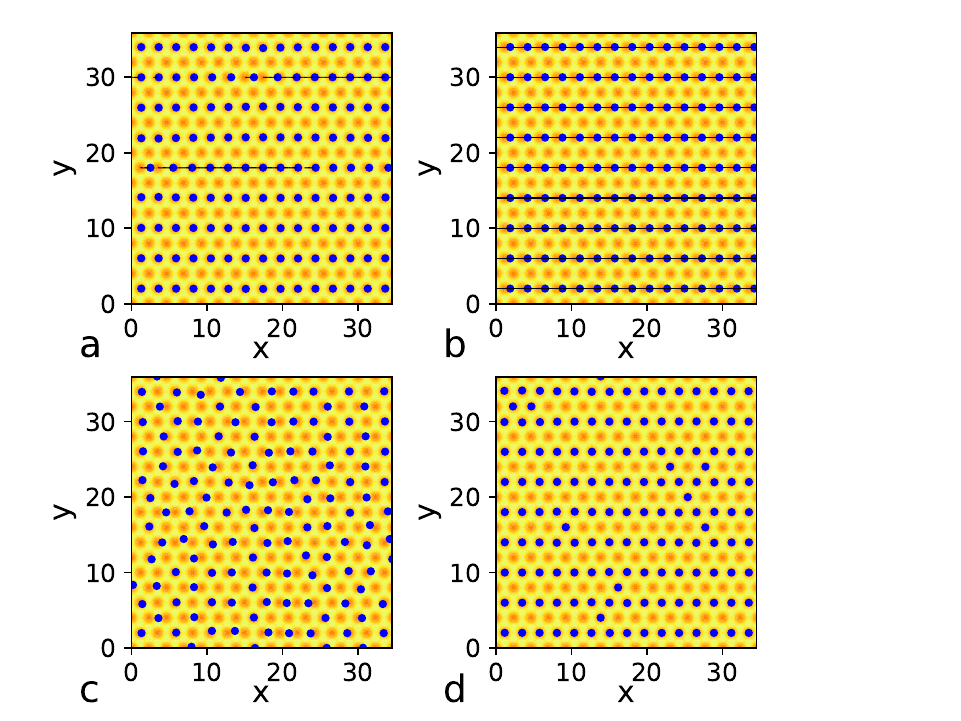}
\caption{The particle locations (blue circles), particle trajectories
(lines), and centers of the hexagonal substrate potential minima (orange
circles) for a system with $T=0$.
(a) Just below $\nu=1/2$ at $F_{D} = 0.04$,
there is anti-kink motion of the two vacancies in the
stripe crystal.
(b) A moving stripe lattice at $\nu = 1/2$ and $F_{D} = 0.09$. 
(c) The moving fluid at $\nu = 0.478$ and $F_{D} = 0.09$;
for clarity, the particle trajectories are not shown.
(d) The zero  bias state at $F_{D} = 0.0$ for $\nu = 0.537$,
showing interstitial particles located between the pinned stripes.
}
\label{fig:7}
\end{figure}

The strong asymmetry in the response
for $\nu < 1/2$ compared to $\nu > 1/2$ is a result of
the formation
of well-defined anti-kinks in the $\nu < 1/2$ state.
To demonstrate this more clearly,
in Fig.~\ref{fig:7}(a), we show the particle locations and trajectories
at $F_{D} = 0.04$ for a system containing
two vacancies in the $\nu = 1/2$ ordered state.
Here, the motion occurs via anti-kinks
that travel opposite to the driving direction
on one-dimensional paths along the chains.
As the drive increases, the anti-kinks move faster, but the number of
anti-kinks
remains constant. 
The motion is similar for $\nu = 0.478$, but more anti-kinks are present.
For $\nu = 1/2$ above depinning, the system forms
a moving stripe lattice, as shown in Fig.~\ref{fig:7}(b) at
$F_{D} = 0.09$.
In contrast,
for $\nu < 1/2$ above the second jump in
$\langle V\rangle$,
instead of a moving stripe there is a moving disordered or fluid state,
as shown in Fig.~\ref{fig:7}(c) for
$\nu = 0.478$ at $F_{D} = 0.09$.
The moving fluid has a higher velocity than the moving
stripe, as illustrated in Fig.~\ref{fig:6}(a) for $F_{D}/F_{p} > 1.0$.
When $\nu > 1/2$, the soliton motion is
lost and the dopants consist of interstitial particles that 
are located between the stripes,
as shown in Fig.~\ref{fig:7}(d)
for $\nu = 0.537$
at $F_{D} = 0.0$.
The $\nu > 1/2$ dopants are more
strongly pinned than the anti-kinks, but when they do move,
they force particles to jump out of the stripes. As a result,
the system depins
into a fluid phase with a high sliding velocity for $\nu>1/2$, instead
of into the moving anti-kink phase with a low sliding velocity found for
$\nu<1/2$.

\begin{figure}
\includegraphics[width=\columnwidth]{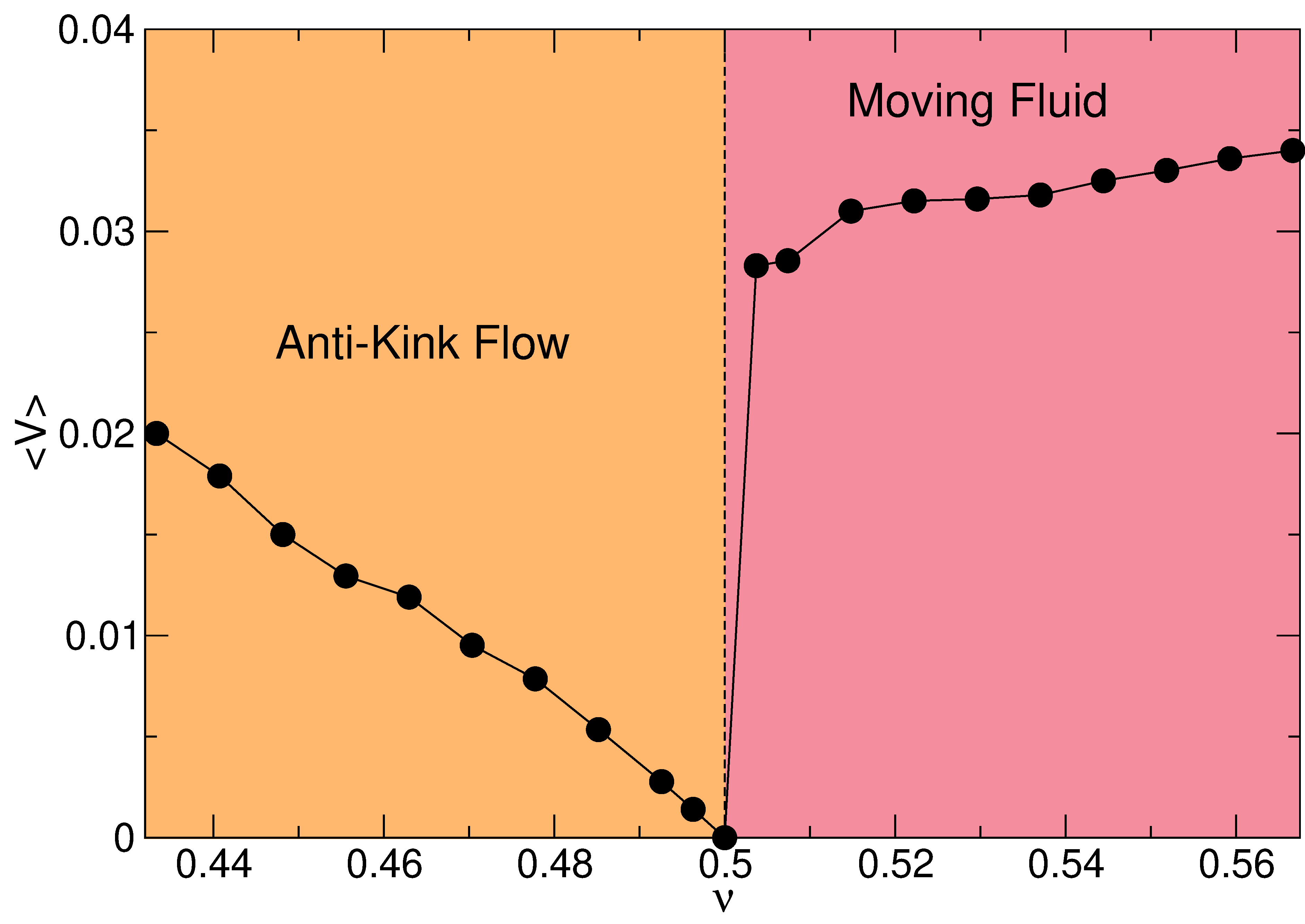}
\caption{$\langle V\rangle$ vs $\nu$
for a sample with $T=0$
at $F_{D} = 0.05$. There is a linear increase in $\langle V\rangle$ with
decreasing $\nu$ for $\nu<1/2$
due to the increase in the number of flowing anti-kinks,
while
a higher velocity moving fluid phase appears for $\nu > 1/2$.
}
\label{fig:8}
\end{figure}

To more clearly demonstrate the asymmetry for $\nu < 1/2$
and $\nu >1/2$, in Fig.~\ref{fig:8} we plot
$\langle V\rangle$ versus $\nu$
in a sample with $T=0$ at $F_{D} = 0.05$.
Here, $\langle V\rangle = 0.0$ at $\nu = 1/2$.
When $\nu < 1/2$, $\langle V\rangle$ increases
linearly with decreasing $\nu$ due to
the increase in the number of anti-kinks
present in the system, which is proportional to the number of
holes added to the stripe crystal.
For $\nu > 1/2$, the velocity is much higher and
depends only weakly on $\nu$.
In this case, additional interstitials located
between the stripes depin,
causing the system to transition into a liquid state with a
high velocity, as illustrated in Fig.~\ref{fig:7}(c).
As the drive is lowered,
the anti-kinks at $\nu<1/2$ continue to flow, but
the $\nu>1/2$ interstitials become pinned.

\begin{figure}
\includegraphics[width=\columnwidth]{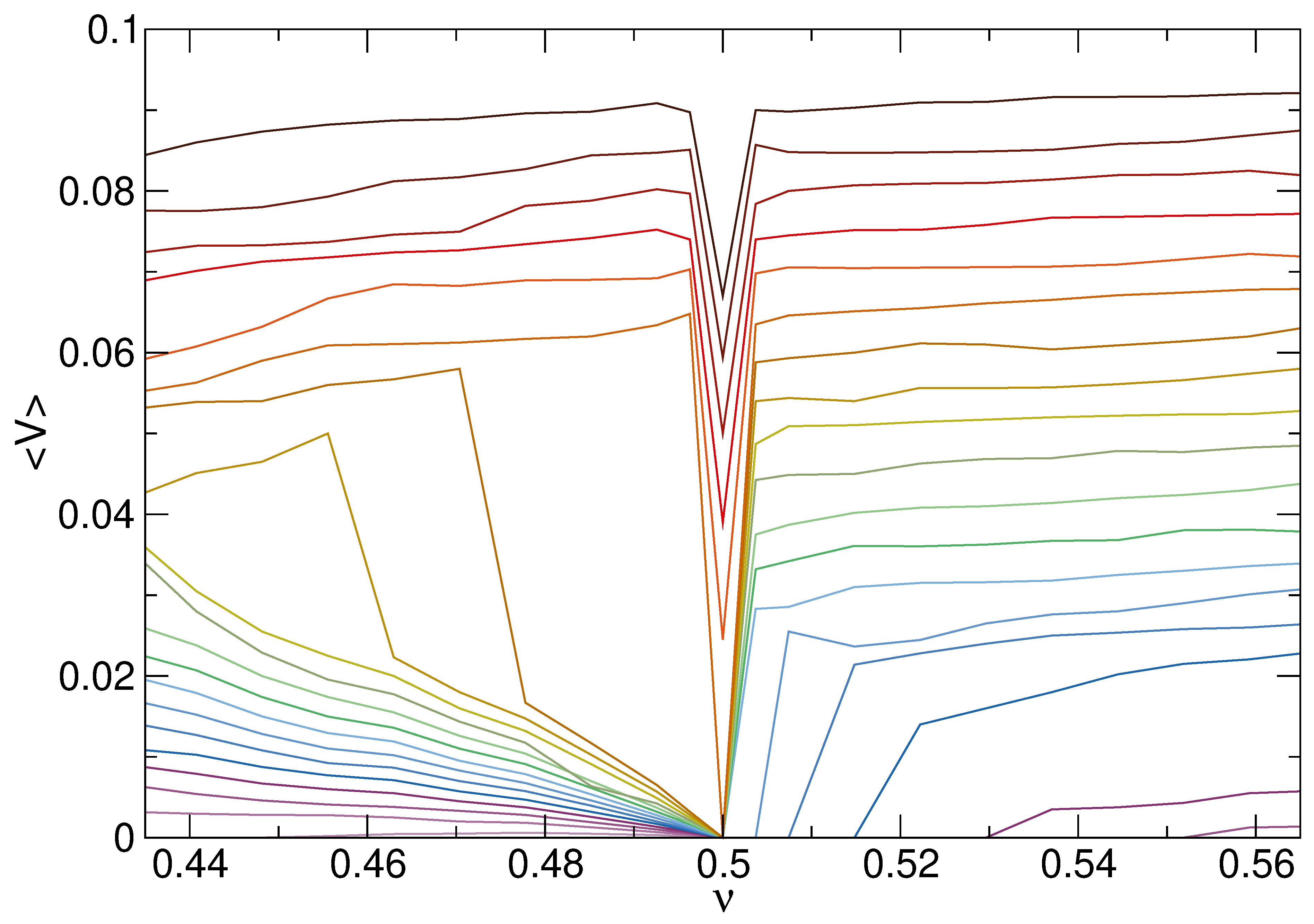}
\caption{$\langle V\rangle$ vs $\nu$
over the range $0.432 < \nu < 0.566$ at $T=0$ and
$F_{D} = 0.0$,
0.005, 0.01, 0.015, 0.02, 0.025, 0.03,
0.035, 0.04, 0.045, 0.05, 0.055, 0.06, 0.065, 0.07, 0.075, 0.08,
0.085, 0.09, 0.095, 0.1, 0.105, and 0.11, from bottom to top.
}
\label{fig:9a}
\end{figure}

In Fig.~\ref{fig:9a} we plot
$\langle V\rangle$ versus $\nu$
over the range $0.432 < \nu < 0.566$
for the system in Figs.~\ref{fig:6} and ~\ref{fig:7}
for drives ranging from $F_{D} = 0.0$ to $F_D=0.11$
in increments of $\delta F_D=0.005$.
Here, the asymmetry in the velocities between $\nu < 1/2$
and $\nu > 1/2$ can be seen more clearly.
For $\nu<1/2$, anti-kink flow occurs from the depinning threshold
up to a maximum drive that increases with increasing $\nu$ from
$F_D=0.065$ at $\nu=0.432$ to $F_D=0.085$ just below $\nu=1/2$.
Lowering the filling in the $\nu<1/2$ regime at a fixed
value of $F_D$ leads to an increase in $\langle V\rangle$ because
this increases the number of anti-kinks present in the system, and
when enough anti-kinks appear, they generate a high velocity fluid-like
state similar to what is observed for $\nu>1/2$.
At $\nu = 1/2$, the system
is pinned up to $F_c = F_{p} = 0.0875$,
and for $F_D>F_c$, 
$\langle V\rangle$ passes through a minimum as the filling is
swept across $\nu=1/2$.

\begin{figure}
\includegraphics[width=\columnwidth]{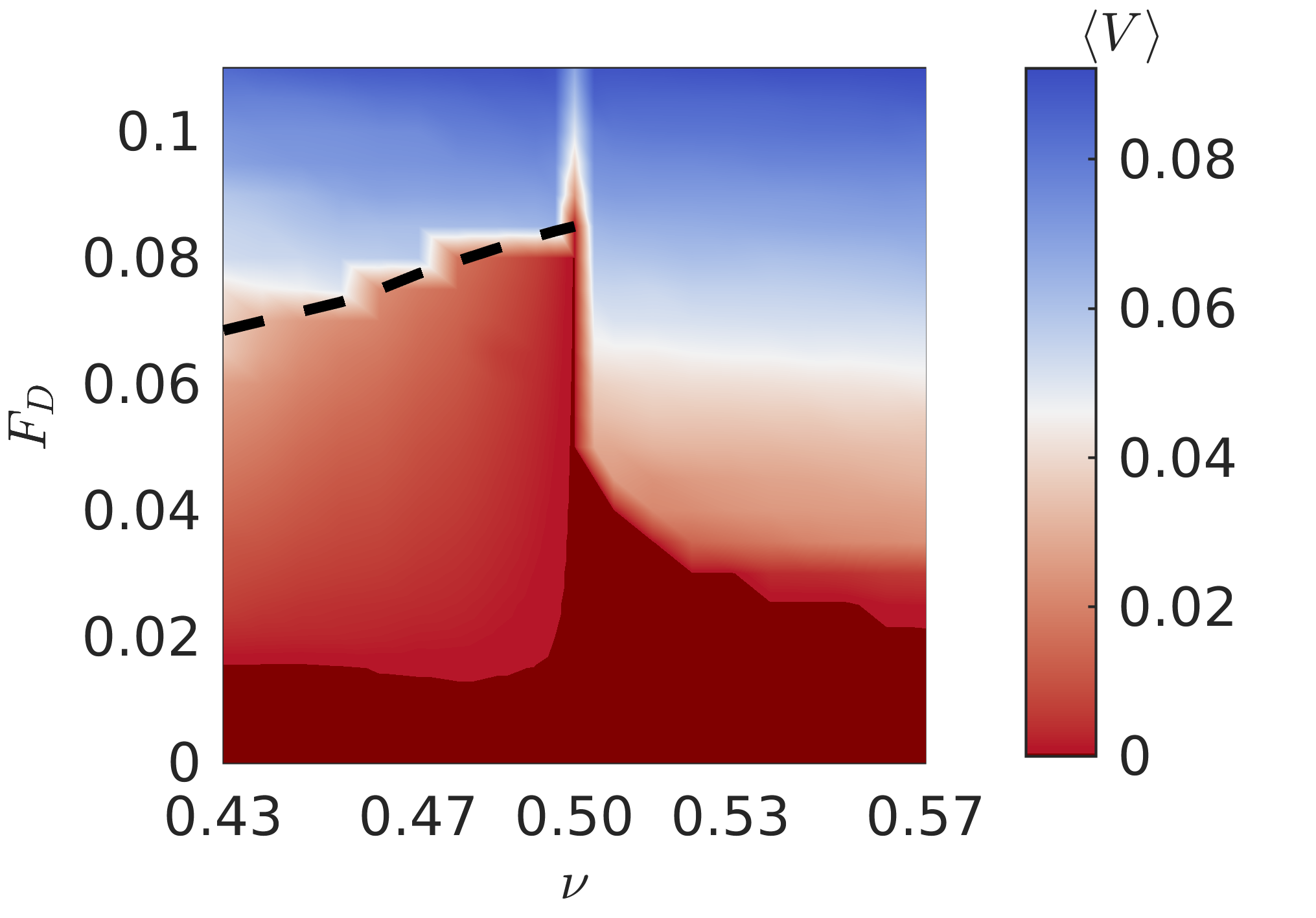}
\caption{A heat map of the velocity $\langle V\rangle$ as a
function of $F_D$ vs $\nu$ constructed using the data
from Fig.~\ref{fig:9a} in a system with $T=0$.
There is an extended region of anti-kink flow for $\nu < 1/2$.
The dashed line denotes the transition from anti-kink to fluid flow.
}
\label{fig:9b}
\end{figure}

Using the data from Fig.~\ref{fig:9a}, in
Fig.~\ref{fig:9b} we construct a heat map of $\langle V\rangle$ as
a function of $F_D$ versus $\nu$.
The dark red region indicates that the depinning threshold is
significantly lower for $\nu<1/2$ than for $\nu \geq 1/2$.
Over the range $0.015 < F_{D} < 0.08$ for $\nu<1/2$,
anti-kink flow is occurring and
the velocity is finite but low.
The dashed line marks
the drive value at which the system transitions from anti-kink flow
to fluid flow. This boundary shifts to lower drives with
decreasing $\nu$ as the density of vacancies increases.
For $\nu>1/2$, the depinning threshold $F_c$ for the interstitial particles
gradually decreases with increasing $\nu$ since the density of interstitial
particles is increasing.  The velocity above depinning is higher in the
$\nu>1/2$ interstitial state than in the $\nu<1/2$ anti-kink flow state
since the interstitials are more mobile than the vacancies.

The heat maps for the region around $\nu=1/3$ in Fig.~\ref{fig:5b}
and for the region around $\nu=1/2$ in Fig.~\ref{fig:9b}
indicate an apparent discontinuity
in the depinning threshold near $\nu = 0.42$.
This arises due to the fact that
we modify $\nu$ by preparing the
system in either the $\nu=1/3$ or $\nu=1/2$ state and doping away from
that state.
For small dopings this procedure gives particle configurations close to
the true ground state, but as the doping becomes heavier, new ground
states with different symmetries arise and the doping protocol does not
capture these states.
If we were to use a different type
of initialization protocol, such as simulated annealing, we expect
to observe additional ordered states
at other higher-order rational filling
fractions such
as $\nu=1/4$, 2/5, 3/5, or $5/6$. These states would be significantly
less robust than the $\nu=1/3$ and $\nu=1/2$ states that are the focus of
the present work.

\begin{figure}
\includegraphics[width=\columnwidth]{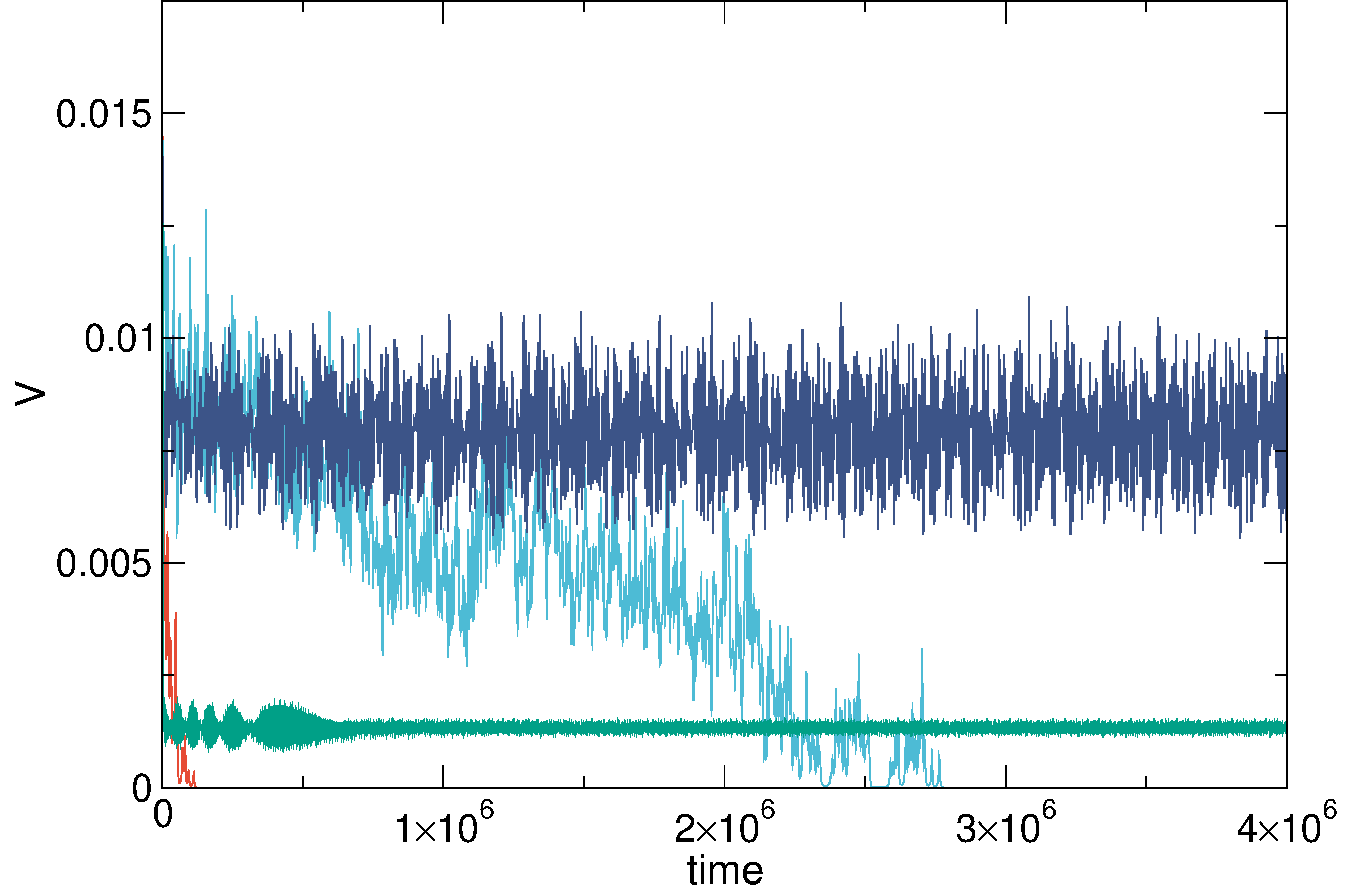}
\caption{Instantaneous velocity $V$ versus time in a sample with $T=0$
at fixed $F_{D} = 0.03$ for
$\nu = 0.39$ (red), 
$\nu = 0.41$ (light blue), $\nu = 0.44$ (dark blue), and $\nu = 0.49$ (green).
}
\label{fig:10}
\end{figure}

If we approach $\nu=0.42$ from above by adding vacancies to the
$\nu=1/2$ state, we find that under application of a driving force,
the system initially depins
into an anti-kink flow state that
transitions into a moving fluid after a short
transient time.
If we dope with even more vacancies so that
$\nu < 0.42$, the moving fluid transitions
to a pinned disordered state after another transient time.
The time scale to reach the pinned
state diverges with increasing $\nu$ up to
a critical filling $\nu_c$,
at which the system remains in a steady state of a
high velocity fluid.
For $\nu>\nu_c$,
the low velocity anti-kink state is stabilized.
To illustrate this, in
Fig.~\ref{fig:10} we plot the instantaneous velocity versus time for
fixed $F_{D} = 0.03$ in a sample with $T=0$.
At $\nu = 0.39$, the velocity drops rapidly to zero and the system enters
a pinned state.
For $\nu = 0.41$, the system is initially
in a fluctuating state, but after a much longer
interval of $2\times 10^6$ simulation time steps, the velocity again
drops to zero and the system reaches a pinned state.
For $\nu = 0.44$, the system remains permanently in a
strongly fluctuating fluid phase with high velocity,
and at $\nu = 0.49$, the system settles into
a sliding anti-kink phase characterized by lower velocity
and reduced velocity fluctuations.

\begin{figure}
    \includegraphics[width=\columnwidth,trim={20 0 90 0},clip]{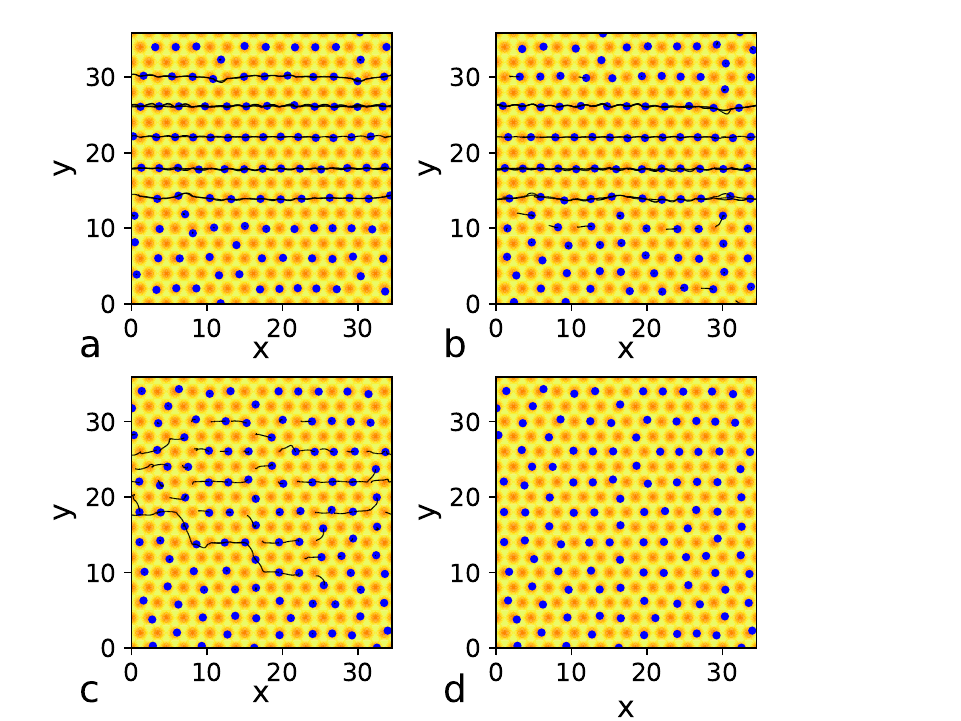}
\caption{The particle locations (blue circles), particle trajectories (lines),
and centers of the hexagonal substrate potential minima (orange circles)
for the system from Fig.~\ref{fig:10} with $T=0$ at $\nu = 0.41$.
(a) The initial phase separated flow.
(b) Reduced flow at later time.
(c) The flow just before reaching the pinned state.
(d) The pinned state.
}
\label{fig:11}
\end{figure}

In Fig.~\ref{fig:11}, we show the particle locations and trajectories
for the system from Fig.~\ref{fig:10} at $\nu = 0.41$ where there is
a long time transient motion before the system reaches a
pinned state. Here, the flow is
initially phase-separated and consists of a large flowing region
surrounded by a pinned region, illustrated in Fig.~\ref{fig:11}(a).
As time progresses, more particles become pinned,
and the trajectories become more meandering, as shown in
Fig.~\ref{fig:11}(b).
The flow gradually decreases and becomes disordered,
as indicated in Fig.~\ref{fig:11}(c).
The final disordered pinned state appears in
Fig.~\ref{fig:11}(d).
The diverging
transient times near $\nu = 0.41$ suggest
that there can be additional glassy-like
behaviors at certain fillings, which could
used to identify disordered charged states in moir{\' e} systems.

\section{2/3 filling}

\begin{figure}
\includegraphics[width=\columnwidth]{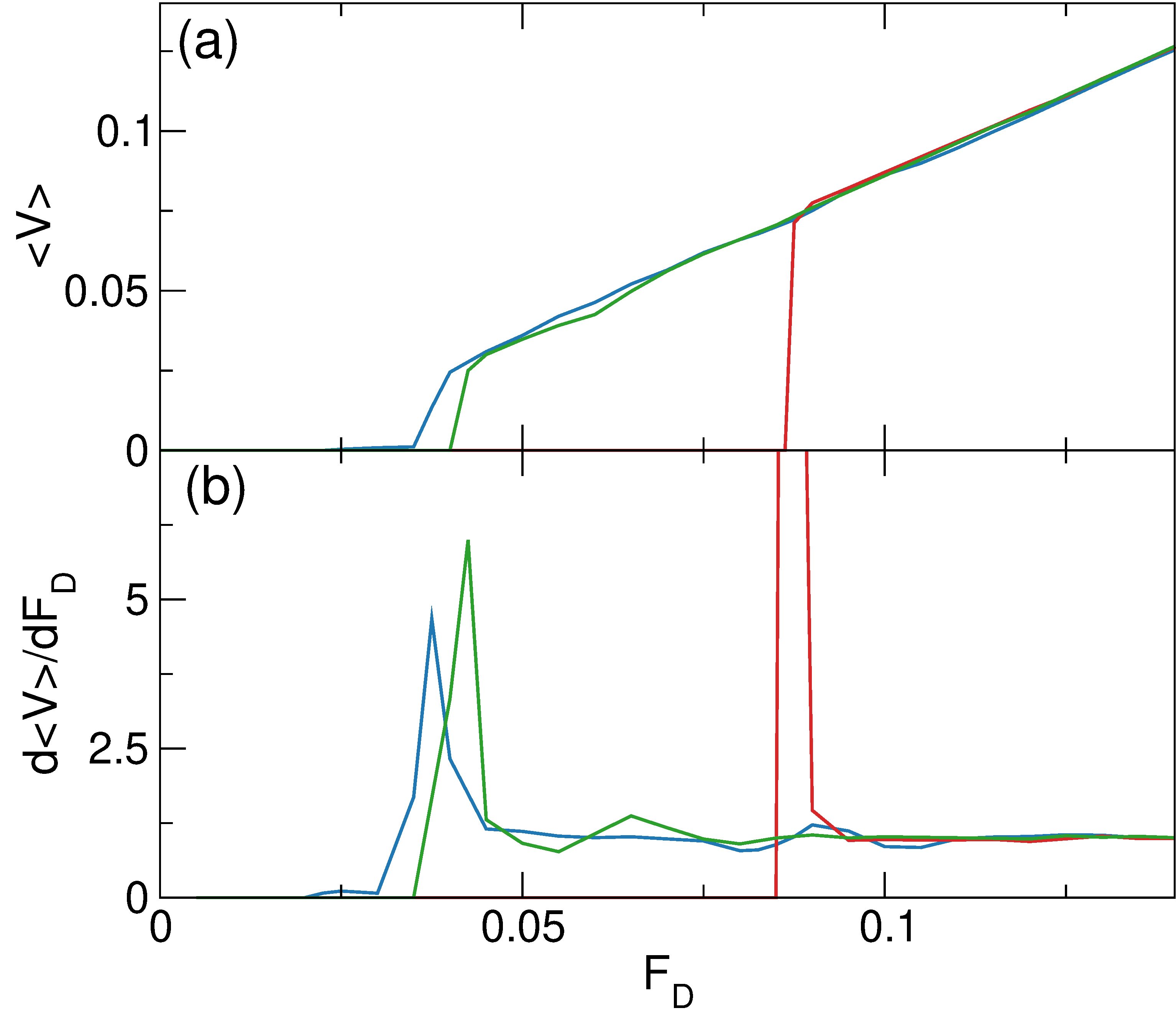}
\caption{$\langle V\rangle$ vs $F_{D}$ for samples with $T=0$
at $\nu = 2/3$ (red), $\nu =0.64$ (blue),
and $\nu = 0.689$ (green). (b) The
corresponding $d\langle V\rangle/dF_{D}$ vs $F_D$ curves.         
}
\label{fig:12}
\end{figure}

\begin{figure}
    \includegraphics[width=\columnwidth,trim={20 0 90 0},clip]{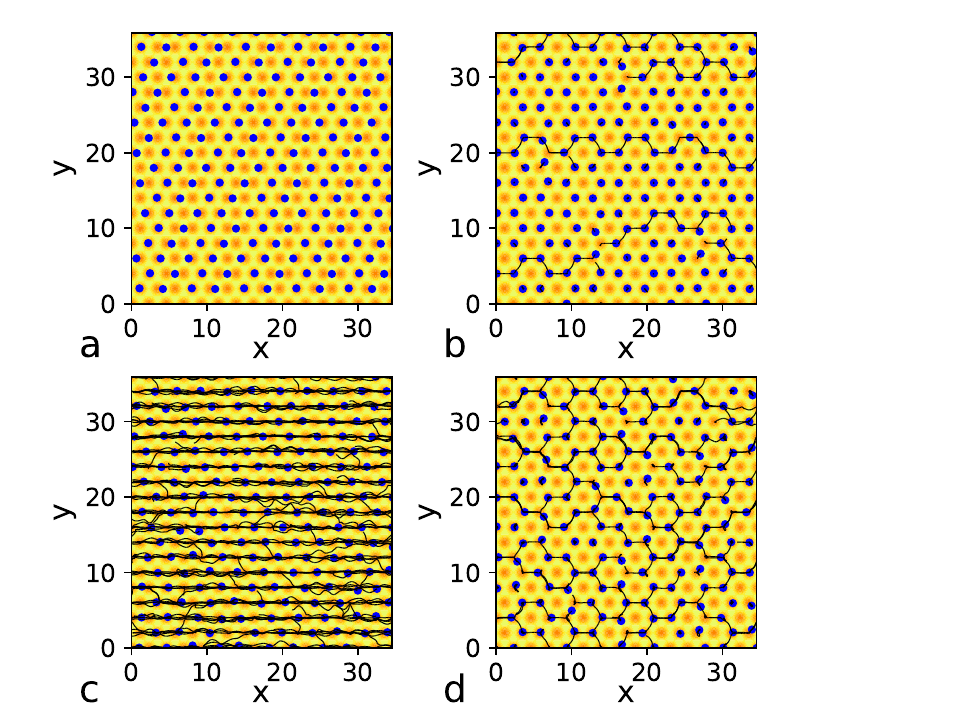}
\caption{The particle locations (blue circles), particle trajectories (lines),
and centers of the hexagonal substrate potential minima (orange circles)
for the system in Fig.~\ref{fig:12} with $T=0$.
(a) A moving floating hexagonal lattice
at $\nu =2/3$ and $F_{D} = 1.0$; for clarity, the trajectories are not shown.
(b) Zig-zag anti-kink flow at $F_{D} = 0.025$ and $\nu = 0.64$.
(c) Fluid flow at $\nu = 0.64$ and  $F_{D} = 0.05$.
(d) An anti-kink flow regime at $\nu = 0.6$ and $F_{D} = 0.025$.
}
\label{fig:13}  
\end{figure}

Next, we consider the dynamics near the $\nu = 2/3$ filling
where the system forms a pinned honeycomb state,
as shown in Fig.~\ref{fig:1}(c).
In Fig.~\ref{fig:12}(a) we plot
$\langle V\rangle$ versus $F_{D}$ at $\nu = 2/3$, $\nu =0.64$, and
$\nu = 0.689$, and in Fig.~\ref{fig:12}(b) we show the
corresponding $d\langle V\rangle/dF_{D}$ versus $F_D$ curves.
At $\nu=2/3$, we find
a discontinuous depinning transition at $F_{D}  = F_p = 0.0875$.
This behavior differs from the continuous depinning
transitions that appear at the $\nu = 1/3$ and $\nu=1/2$ fillings.
For the $\nu = 1/3$ and $\nu=1/2$ fillings, the moving states
have the same symmetry as the pinned states.
In contrast, at $\nu = 2/3$, 
the system does not form a moving honeycomb
lattice with the same symmetry as the pinned state but
instead adopts
a moving hexagonal lattice configuration, as shown in Fig.~\ref{fig:13}(a)
at $F_{D} = 1.0$.
Here, the hexagonal
lattice does not couple strongly to the substrate
and moves faster than the fluid like state that forms for the same value
of $F_D$ at the incommensurate fillings.
The $\nu=2/3$ system forms
a moving floating solid state that is distinct from the
commensurate moving solid states found at $\nu = 1/3$ and $\nu=1/2$.
At
$\nu = 0.64$
in Fig.~\ref{fig:12},
there is
a window
spanning the range $0.02 < F_D < 0.034$ where anti-kink flow occurs and
the velocity is low but finite.
Unlike the one-dimensional anti-kink motion found for $\nu=1/2$, here
the anti-kinks follow a zig-zag pattern
as shown in Fig.~\ref{fig:13}(b) at $F_{D} = 0.025$.
For higher drives,
the anti-kink flow breaks up
the background lattice, and the systems enters
a disordered higher velocity
fluid flow state, as illustrated in Fig.~\ref{fig:13}(c)
at $F_{D} = 0.05$.
The transition from the anti-kink flow to the
moving fluid is accompanied by 
a large jump
in $\langle V\rangle$
that appears near $F_{D} = 0.04$.
For $\nu = 0.689$, the system depins directly into a fluid flow state
similar to the flow illustrated in Fig.~\ref{fig:13}(c).
Between $\nu=2/3$ and $\nu=0.689$,
the soliton or anti-kink flow can persist,
and Fig.~\ref{fig:13}(d) shows the
intertwined channel structure of the trajectories in the
anti-kink flow regime at $\nu = 0.6$.
Here, particles can easily jump from lane to lane,
so in addition to the motion along the driving direction,
there is a long time diffusion
of the particles in the direction transverse to the drive.

\begin{figure}
\includegraphics[width=\columnwidth]{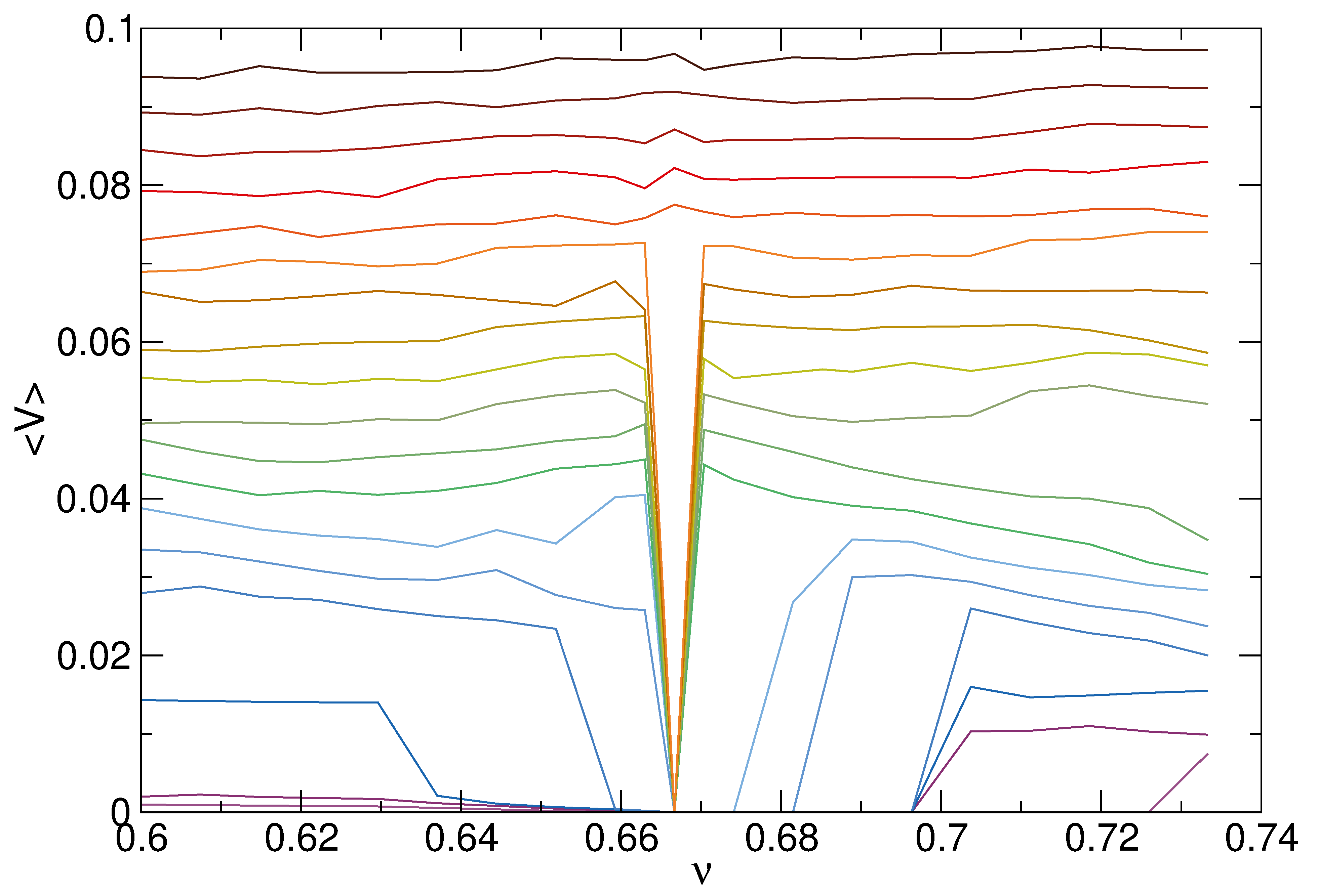}
\caption{$\langle V\rangle$ vs $\nu$ over the range $0.6 < \nu < 0.73$ for
a system with $T=0$ at
$F_{D} = 0.0$,
0.005, 0.01, 0.015, 0.02, 0.025, 0.03,
0.035, 0.04, 0.045, 0.05, 0.055, 0.06, 0.065, 0.07, 0.075, 0.08,
0.085, 0.09, 0.095, 0.1, 0.105, and 0.11, from bottom to top.
}
\label{fig:14a}  
\end{figure}

\begin{figure}
\includegraphics[width=\columnwidth]{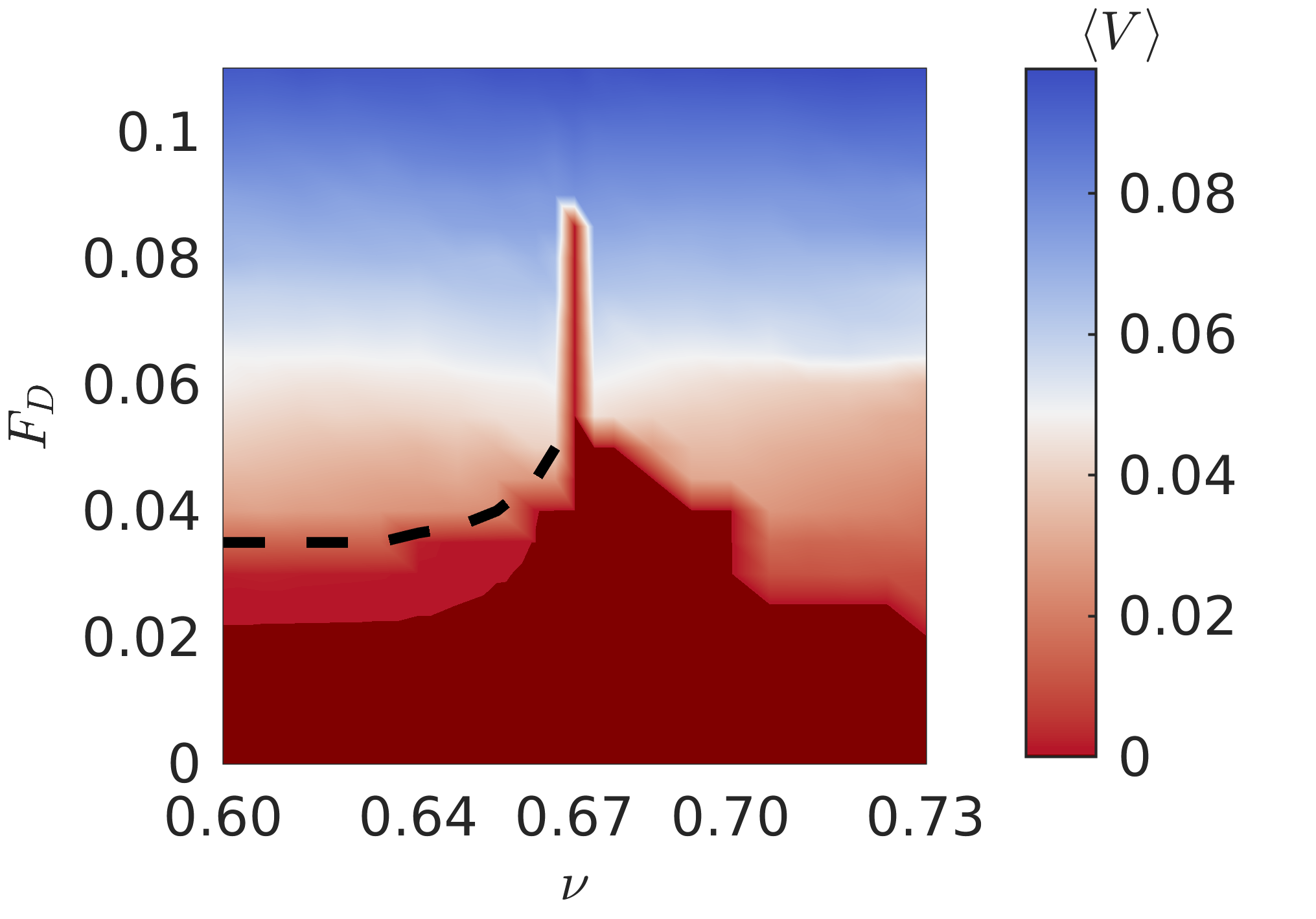}
\caption{A heat map of the velocity $\langle V\rangle$ as a function
of $F_D$ vs $\nu$ constructed  using the data from Fig.~\ref{fig:14a}.
The dark red region indicates
the pinned phase.
The dashed line indicates the boundary between the anti-kink flow and
moving fluid states
for $\nu < 2/3$.
}
\label{fig:14b}  
\end{figure}

In Fig.~\ref{fig:14a} we plot
$\langle V\rangle$
versus $\nu$
over the range $0.6 < \nu < 0.73$
for $F_{D} = 0.0$ to
$0.11$ in drive increments of $\delta F_D=0.005$.
For $F_{D} < 0.0875$,
$\langle V\rangle = 0.0$ at $\nu = 2/3$, while for
$F_{D} \geq 0.0875$, the
velocity at $\nu = 2/3$ passes through 
a small peak, which is
the opposite of the behavior found for $\nu = 1/3$ and $\nu=1/2$,
where the velocity was diminished at the commensurate filling
for high drives above depinning.
For $0.02 < F_{D} < 0.045$, there is a regime
of anti-kink flow
for $\nu < 2/3$.
The depinning threshold
shifts to higher drives immediately above $\nu > 2/3$
since the additional dopant particles move to the
center of the ordered structures in the $\nu = 2/3$ honeycomb arrangement.
Figure~\ref{fig:14b} shows a heat map
of $\langle V\rangle$ as a function of $F_D$ versus $\nu$
for the system in Fig.~\ref{fig:14a} indicating the locations of the
pinned and moving phases.
The dashed line marks
the boundary between the anti-kink flow and fluid flow states
for $\nu < 2/3$.  

At $\nu = 1.0$, we find that the depinning
transition is continuous and
the system depins from the hexagonal pinned lattice shown in
Fig.~\ref{fig:1}(d) into a moving hexagonal solid.
There is also a regime of anti-kink flow for $\nu < 1.0$.
We will describe the dynamics near $\nu = 1.0$ in more detail elsewhere
since for $\nu > 1.0$, the substrate minima become doubly occupied,
producing entirely distinctive types of flows compared to
systems with $\nu \leq 1.0$.

\section{Drive Dependent Melting}

We next consider the effect of thermal fluctuations
and melting on the depinning dynamics.
We focus on the $\nu = 1/3$ filling, but
similar behavior appears
at the higher fillings.
The melting transition at $\nu=1/3$ can be characterized by
the onset of random thermal hopping of charges, which causes the
configuration to become disordered.
We note that in previous work, it was shown that the melting temperature is
larger at commensurate fillings compared to incommensurate fillings
by a factor of two or more \cite{Matty22}.

\begin{figure}
\includegraphics[width=\columnwidth]{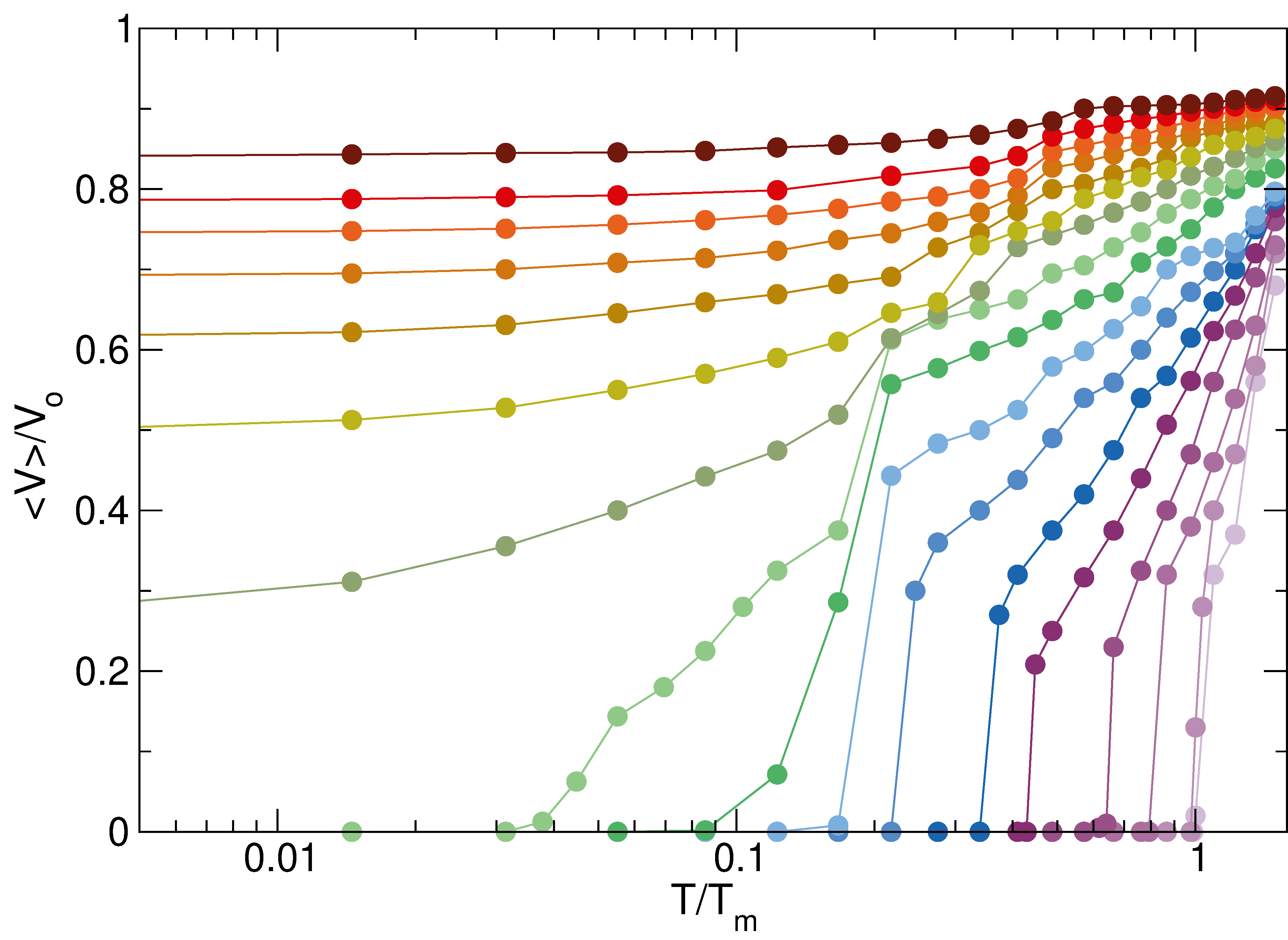}
\caption{$\langle V\rangle/V_{0}$ vs
$T/T_{m}$
for the system from Fig.~\ref{fig:1}(a) 
at $\nu=1/3$ under drives of
$F_{D}= 0.0025$ to $0.16$ from bottom to top.
Here $V_{0}$ is the substrate-free velocity and
$T_m$ is the melting temperature for zero drive. 
}
\label{fig:15a}
\end{figure}

In Fig.~\ref{fig:15a} we plot
$\langle V\rangle/V_{0}$ versus $T/T_{m}$
for drives spanning the range $F_{D}=0.0025$ to $F_D=0.16$ in the sample
from Fig.~\ref{fig:1}(a)
at $\nu=1/3$.
Here $V_{0}$ is the substrate-free velocity
and $T_m$ is the melting temperature in the presence of the substrate
but in the absence of a drive.
For the smallest drive we consider, $F_{D}= 0.0025$, we find that
the velocity becomes finite when the system melts.
In general, for $T/T_{m} > 0.35$ all of the sliding states
are fluid like;
however, for $T/T_{m} < 0.35$, it is possible
for the system to depin directly into a moving crystal or
to order dynamically into a moving crystal state
at high enough drives.
For $F_{D} > 0.0875 = F_p$,
the velocities are always finite,
while at lower drives, the velocities become
finite only when the temperature is increased to a sufficiently high
level.
The $\nu = 1/3$ system
can show
finite transport for $T/T_m > 0.65$ under drives that are nearly
five times smaller
than the zero temperature depinning threshold.

\begin{figure}
\includegraphics[width=\columnwidth]{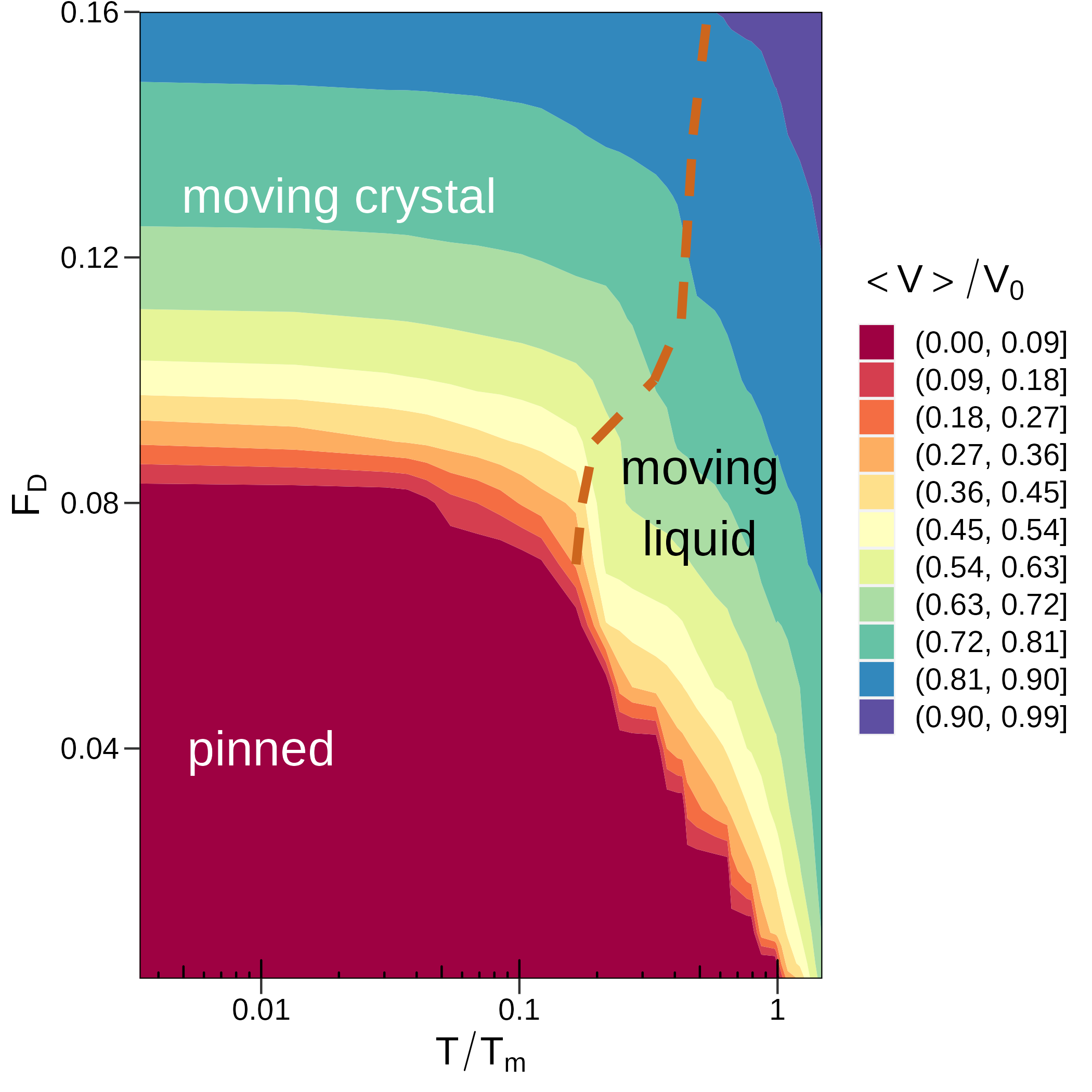}
\caption{A heat map of 
  $\langle V\rangle/V_{0}$ as a function of $F_D$ vs $T/T_m$ constructed
  from the data in Fig.~\ref{fig:15a} for a sample with $\nu=1/3$.
  The dashed line indicates the boundary between
  the moving crystal and moving fluid states.
}
\label{fig:15b}
\end{figure}

In Fig.~\ref{fig:15b}, we plot a heat map of $\langle V\rangle/V_0$
as a function of $F_D$ versus $T/T_m$
for the $\nu=1/3$ system from Fig.~\ref{fig:15a},
where we highlight the locations of
the pinned and moving phases. The dashed line indicates the
separation between the moving crystal and moving liquid states.
For $T/T_m>0.1$, we find
that the depinning temperature
is strongly reduced with increasing drive.
Additionally, the dynamic reordering transition from a moving
liquid to a moving crystal shifts to higher drives as the
temperature increases.
Similar behavior of the dynamic reordering transition has been observed
for superconducting vortices and colloidal particles driven over
disordered substrates,
where the system can depin into a fluid and then reorder under driving
into a crystalline state,
but the drive needed to reorder the system diverges
as the substrate-free melting temperature is approached
\cite{Reichhardt17}.
The results in Fig.~\ref{fig:15b} suggest that the depinning
threshold at commensurate fillings
should be accessible experimentally even
for low drives if finite temperatures are applied.
We find similar behaviors at $\nu = 1/2$ and $\nu = 2/3$, except that
the thermal depinning thresholds decrease even more rapidly
with increasing driving force
than for the $\nu=1/3$ filling.

\begin{figure}
    \includegraphics[width=\columnwidth,trim={20 0 90 0},clip]{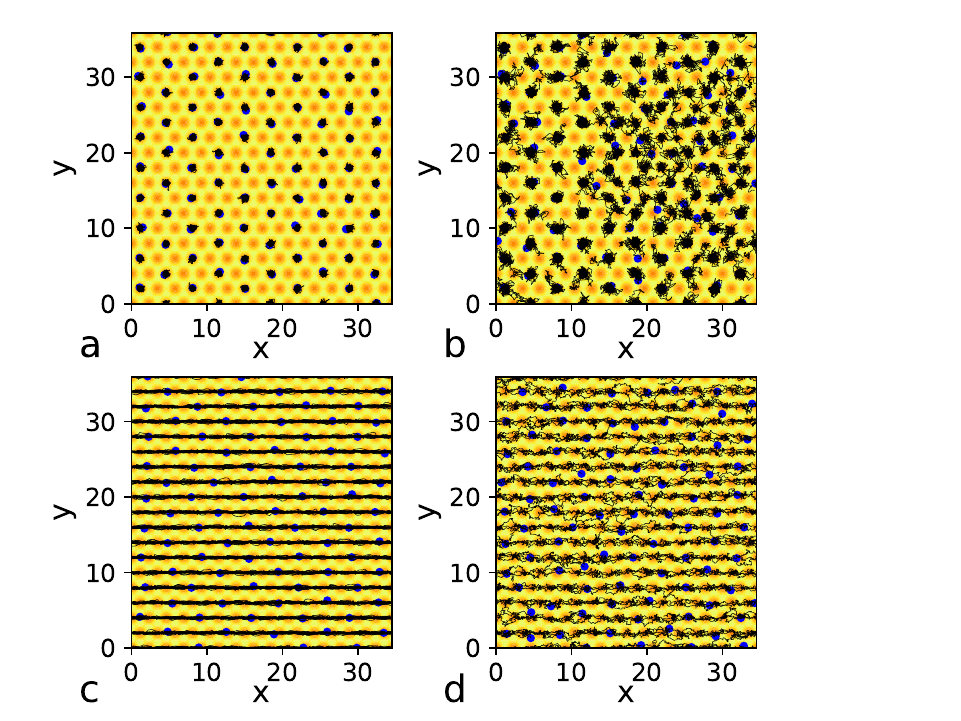}
\caption{The particle locations (blue circles), particle
trajectories (lines), and centers of the hexagonal substrate potential
minima (orange circles) for a
finite temperature system at $\nu=1/3$.
(a) A pinned state at $T/T_{m} = 0.48$ and $F_D=0$.
(b) Thermally induced hopping for 
$T/T_m = 1.1$ and $F_D=0$.
(c) A moving crystal phase at
$T/T_m = 0.12$ and
$F_{D} = 0.11$.
(d) A  moving fluid 
at $T/T_m = 0.67$ and $F_D=0.11$.
}
\label{fig:16}
\end{figure}

Figure~\ref{fig:16}(a) shows the particle locations and trajectories
for zero bias, $F_D=0$, at $T/T_{m} = 0.48$ and $\nu = 1/3$,
where some thermally induced motion occurs
but the particles remain pinned.
In Fig.~\ref{fig:16}(b), the same system at
$T/T_m = 1.1$ exhibits hopping throughout the sample,
and the hexagonal ordering of the particles is lost.
Figure~\ref{fig:16}(c) shows the moving crystal phase at
$F_{D} = 0.11$ and
$T/T_m = 0.12$, where the particles form a moving hexagonal lattice
and there is some
smearing of the trajectories due to the thermal motion.
If the temperature is increased to
$T/T_m = 0.67$, where the system is disordered,
there is a moving fluid state as illustrated in
Fig.~\ref{fig:16}(d).

\begin{figure}
    \includegraphics[width=\columnwidth,trim={20 0 90 0},clip]{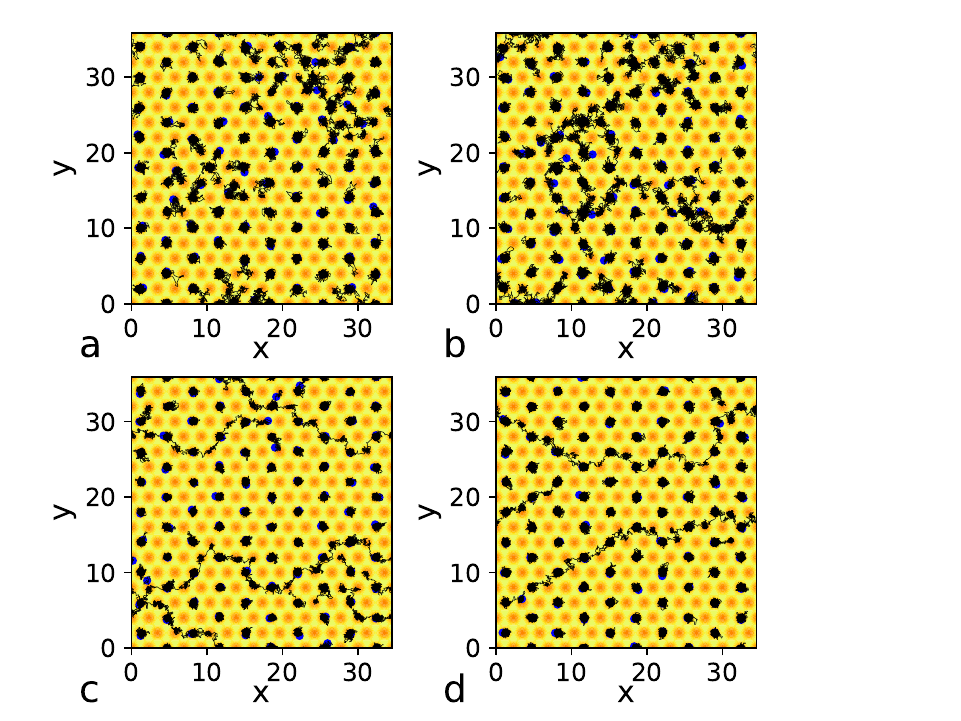}
\caption{The particle locations (blue circles), particle trajectories (lines),
and centers of the hexagonal substrate potential minima (orange circles).
(a) A zero bias system with $F_{D} = 0.0$ at $T/T_m = 0.48$ and $\nu = 0.31$.
(b) The same as in (a) but for $\nu = 0.36$.
(c) A finite bias state
at $T/T_m = 0.4867$ and $F_{D} = 0.02$
for a $\nu = 1/3$ sample where two dopant
particles have been added.
(d) The same as in (c) but with two vacancies.
}
\label{fig:17}
\end{figure}

The thermal effects are stronger at the incommensurate fields.
In Fig.~\ref{fig:17}(a), we show the particle positions
and trajectories for the zero bias
$\nu=0.31$ system at $T/T_m = 0.48$, a temperature for which
the $\nu = 1/3$ state was still pinned as indicated
in Fig.~\ref{fig:16}(a).
Due to the motion of vacancies in the $\nu=0.31$ system,
there is considerable thermal hopping in Fig.~\ref{fig:17}(a),
and at $\nu=0.36$ in Fig.~\ref{fig:17}(b),
there is even more motion than for $\nu < 1/3$.
This suggests that at finite temperatures, the incommensurate states can
be viewed as a combination of
a pinned solid with local fluid-like regions. 

In Fig.~\ref{fig:17}(c), we show the trajectories at $T/T_m=0.4867$ and
$F_D=0.02$ in a 
$\nu = 1/3$ system that has had two additional dopant particles added
to it.
Localized motion occurs
consisting of interstitials that travel on average in the direction of
the drive.
The motion is not continuous but consists of hopping from one well to another.
Figure~\ref{fig:17}(d) illustrates the $\nu=1/3$ system where two particles
have been removed to create two vacancies. Here
the same type of localized motion of the vacancies appears.
Close to and on either side of the $\nu=1/3$ filling,
the holes or interstitials
remain pinned for lower drives or lower temperatures.
We have also observed similar thermal creep motion of
interstitials or vacancies
at incommensurate states close to the higher
commensurate fillings.

\begin{figure}
\includegraphics[width=\columnwidth]{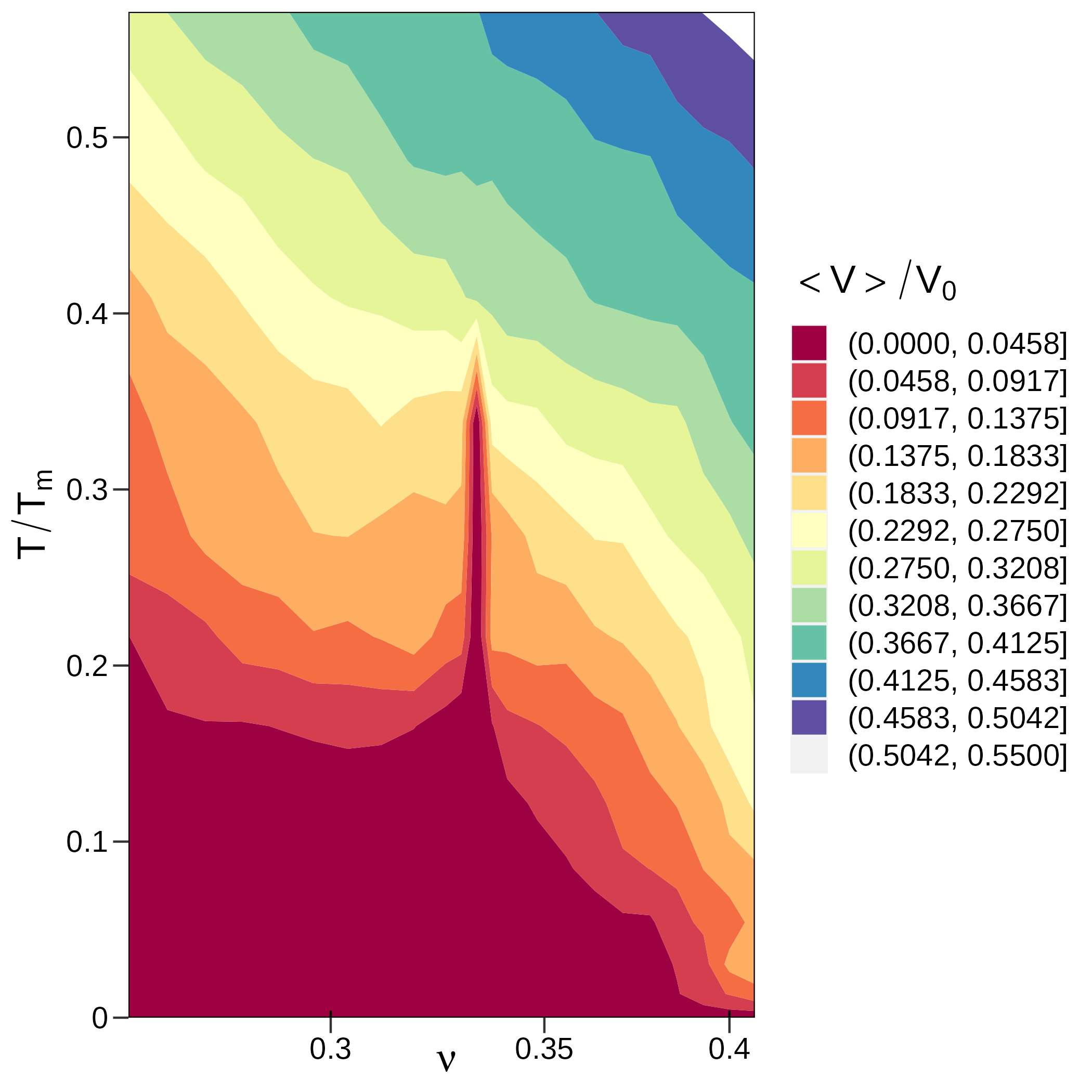}
\caption{A heat map of $\langle V\rangle/V_0$ as a function of
$T/T_m$ vs $\nu$ over the range $0.26 < \nu < 0.41$ at fixed $F_{D} = 0.4$.
}
\label{fig:18}
\end{figure}

In Fig.~\ref{fig:18}, we show a heat map of $\langle V\rangle/V_0$ as a
function of $T/T_m$ versus $\nu$ over
the range $0.26 < \nu < 0.41$ at fixed $F_{D} = 0.4$.
Here, it can be seen that depinning
occurs near $T/T_m= 0.4$ at $\nu = 1/3$,
near $T/T_m = 0.2$ for $\nu < 1/3$,
and close to zero temperature for $\nu > 0.4$.

\section{Discussion}

Our results could be confirmed by
performing transport experiments with external drives to look for
nonlinear features in the transport curves and changes
in the conduction threshold similar to those found
in other systems where Wigner crystals occur.
Additional measurable effects include
the conduction noise, which should have very distinct spectra
in the moving crystal, moving fluid, and anti-kink flow states.
At the
incommensurate fillings, we would expect more glassy-like behavior
to appear due to the
disordered nature of the states \cite{Tan23}.
If ac driving were applied, for
smaller
ac drive amplitudes the commensurate filings should be ordered enough that
the changes never leave the pinning
wells; however, the solitons at incommensurate fillings
could flow much further or
undergo well-to-well hopping.
Another possible experimental probe would
be to inject or move charges with a localized tip;
this process should require much lower energies
at incommensurate
fillings than at
commensurate fillings.
It is also likely that
the depinning and transport properties would be anisotropic.
For example, at $\nu=1/2$,
the soliton flow runs along the stripes, but the response could be
significantly modified
if the drive were applied perpendicular to the stripes.

In our study, we held the substrate strength fixed,
but this may be tunable
experimentally by changing the coupling between
layers, where weaker coupling would correspond to
weaker substrates.
Modifying the substrate strength could
make it possible to realize a
transition to a floating Wigner solid that is
similar to the Aubry transitions observed
for charged colloidal particles on periodic substrates \cite{Brazda18}.
We have only focused on fillings $\nu < 1.0$,
but we expect that a variety of other dynamics could
arise at higher filings such as
$\nu = 2$, $\nu=3$, or higher values
where Wigner molecules have been predicted to exist and
ordered clusters can appear \cite{Li23,Reddy23,Kometter23}.
It would also be interesting to see if the conduction curves
at finite temperatures
show
distinct signatures
at commensurate versus incommensurate fillings that could indicate
the presence of glassy-like behavior at incommensurate fields
similar to what has been found in
superconducting vortex glasses \cite{Nattermann00}.
The ordering of the charges at commensurate states
has many similarities to
superconducting vortex
ordering in wire network arrays \cite{Chang92},
superconductors with triangular periodic pinning arrays
\cite{Baert95,Reichhardt01a,Grigorenko03},
and charge ordering on square
arrays \cite{Rademaker13},
so it is possible
that effects observed those systems
could also arise for charge ordering in the moir{\' e} system.

Our results are obtained in the classical limit,
and the ordered states we find are in
good agreement
with classical Monte Carlo \cite{Matty22}
and Hartree-Fock calculations \cite{Ung23};
however, in the
work by Ung {\it et al.} \cite{Ung23}, the Hartree-Fock calculations
showed that the energy landscapes can be complex, with several competing
states that could, in some cases, lead to disordered configurations.
Such effects could be amplified at finite
drives or temperatures.
Another effect that we did not consider in our model is
magnetic degrees of freedom that could create additional
coupling leading to the formation of dimer-like states
\cite{Duran23,Song23,Kaushal22}.
Polaronic effects \cite{Arsenault24},
applied diamagnetic fields, or additional random
disorder in the material could
also increase or decrease the pinning thresholds compared to what we observe.

Our work is focused on modeling
the dynamics of Wigner crystals in moir{\' e} heterostructures;
however, given the recent observations
of Wigner crystals in graphene systems \cite{Tsui24},
another approach would be to imprint some form of 
external hexagonal substrate on the graphene or
to couple the graphene to
another system that hosts Wigner crystals
in order to create a hexagonal substrate. 

If sliding motion in the form of crystals or kinks could be realized
in moir{\' e} heterostructures, it would
open a wealth of possible functionalities for this class of system.
The generalized Wigner crystals would also represent
a new
class of incommensurate-commensurate transitions
where long-range interactions are important.
In previous commensuration-incommensuration work
with colloidal particles,
superconducting vortices, and frictional systems,
the interactions often extended only to nearest neighbors
or at most to the first few neighbors. Long range interactions can produce
distinctive behaviors compared to the short range interactions.
In our work, we focused on the case where the particles were strongly
interacting with each other,
but it would also be interesting to 
to study the pinning and sliding
behavior for varied levels of screening of the interactions.

\section{Summary} 

We have numerically investigated the pinning, 
sliding and melting of
generalized Wigner crystal states in moir{\' e} heterostructures
for commensurate fillings of
$\nu=1/3$, 1/2, and $2/3$ as well dopings above and below these fillings.
At $\nu=1/3$, the charges form a pinned hexagonal lattice.
The depinning threshold reaches its maximum value at $\nu = 1/3$, where
there is a single sharp depinning transition
into a moving hexagonal crystal at low temperatures.
For fillings just above or below the commensurate fillings,
the depinning threshold is reduced, and multiple depinning transitions
may appear.
At $\nu = 1/2$, the system forms a stripe crystal,
and for dopings of $\nu < 1/2$, the depinning threshold is strongly reduced
and the depinning occurs via the motion of anti-kinks along the
stripes.
For dopings above $\nu=1/2$,
the additional particles form interstitial inclusions between the
stripes and the depinning threshold is increased compared to doping with
an equivalent number of vacancies,
but when 
depinning occurs, the system enters
a high velocity fluid state.
For $\nu = 2/3$, the system forms a honeycomb lattice
that shows a sharp, discontinuous
depinning transition to a floating hexagonal lattice.
For $\nu < 2/3$, we find a different type of
anti-kink sliding phase where the motion follows localized zig-zag patterns.

For finite temperatures and no driving,
the $\nu=1/3$ states have a well defined melting temperature $T_m$ above
which the particles begin to
hop out of the wells.
The effective melting temperature, which can be measured by detecting the
onset of conduction, is reduced at finite drives. For
higher drives, the system transitions from a pinned hexagonal
lattice to a moving fluid as the temperature increases,
while at lower drives, the
system depins into a moving hexagonal lattice.
At commensurate 
fillings, introduction of thermal fluctuations causes a dramatic drop
in the depinning threshold, and the motion above depinning can take the
form of localized hopping of interstitials or vacancies through the
lattice.
We show that the sliding phases can be detected via
features in the transport and differential transport curves,
and that these phases should be readily accessible experimentally,
particularly for incommensurate fillings and finite temperatures.
Realizing sliding anti-kink or crystalline states
would open up a variety of new functionalities for
moir{\' e} systems. The dynamics of generalized Wigner
crystals in these systems
represent a new class of
commensurate-incommensurate
transitions with long-range interactions.

\smallskip

\begin{acknowledgments}
We thank C.D. Schimming for helpful discussions.
We gratefully acknowledge the support of the U.S. Department of
Energy through the LANL/LDRD program for this work.
This work was supported by the US Department of Energy through
the Los Alamos National Laboratory.  Los Alamos National Laboratory is
operated by Triad National Security, LLC, for the National Nuclear Security
Administration of the U. S. Department of Energy (Contract No. 892333218NCA000001).
\end{acknowledgments}

\bibliography{mybib}

\end{document}